\documentclass[11pt]{article}
\usepackage{epsfig}
\usepackage{amsfonts}
\usepackage{amsmath}
\usepackage{amssymb}

\newcommand{\w}{\omega}

\newcommand{\slashPsub}{\slash\hspace{-0.48em}P}

\newcommand{\Nb}{\bar N}
\newcommand{\Fp}{F_\pi}

\newcommand{\mpi}{m_{\pi}}

\newcommand{\dslash}[1]{#1 \llap{/\kern-0.5pt}}
\newcommand{\Dslash}[1]{#1 \llap{/\kern+1.2pt}}
\newcommand{\DDslash}[1]{#1 \llap{/\kern+2.3pt}}
\newcommand{\dslashh}[1]{#1 \llap{/\kern+1pt}}

\newcommand{\bea}{\begin{eqnarray}}
\newcommand{\eea}{\end{eqnarray}}
\newcommand{\bma}{\begin{pmatrix}}
\newcommand{\ema}{\end{pmatrix}}
\newcommand{\nn}{\nonumber}

\begin{document}
\begin{titlepage}

\vspace{2.0cm}

\begin{center}
{\Large\bf 
Parity violation in proton-proton scattering\\[0.3em] from chiral effective field theory}

\vspace{1.7cm}

{\large \bf   J. de Vries$^{1}$, Ulf-G. Mei{\ss}ner$^{1,2}$, E. Epelbaum$^{3}$, N. Kaiser$^{4}$} 

\vspace{0.5cm}

{\large 
$^1$ 
{\it Institute for Advanced Simulation, Institut f\"ur Kernphysik, 
and J\"ulich Center for Hadron Physics, Forschungszentrum J\"ulich, 
D-52425 J\"ulich, Germany}}

\vspace{0.25cm}
{\large 
$^2$ 
{\it Helmholtz-Institut f\"ur Strahlen- und Kernphysik and Bethe Center for 
Theoretical Physics, Universit\"at Bonn, D-53115 Bonn, Germany}}

\vspace{0.25cm}
{\large 
$^3$ 
{\it Institut f\"ur Theoretische Physik II, Ruhr-Universit\"at Bochum, 44780 Bochum, Germany}}

\vspace{0.25cm}
{\large 
$^4$ 
{\it Physik Department T39, Technische Universit\"at M\"unchen, D-85747 Garching, Germany}}

\end{center}

\vspace{1.5cm}

\begin{abstract}
We present a calculation of the parity-violating longitudinal asymmetry in 
proton-proton scattering. The calculation is performed in the framework of
chiral effective field theory which is applied systematically to both the  
parity-conserving and parity-violating interactions. The
asymmetry is calculated up to next-to-leading order in the parity-odd nucleon-nucleon
potential. At this order the asymmetry depends on two parity-violating 
low-energy constants: the weak pion-nucleon coupling constant $h_\pi$ and one 
four-nucleon contact coupling. By comparison with the existing data, 
we obtain a rather large range for $h_\pi= (1.1\pm
2 )\cdot 10^{-6}$. This range is consistent with theoretical estimations and 
experimental limits, but more data are needed to pin down a better constrained value. 
We conclude that an additional measurement of the asymmetry around $125$~MeV lab energy
would be beneficial to achieve this goal.
\end{abstract}

\vfill
\end{titlepage}

\section{Introduction}
The observation of parity ($P$) violation in the weak interaction is one of
the pillars on which the Standard Model of particle physics was built. 
In the Standard Model 
parity violation is implemented by specifying different gauge-symmetry representations of the chiral fermions which has the consequence that only left-handed quarks and leptons 
participate in the (charged current) weak interaction. At the fundamental level, parity violation originates from the exchange of the charged (and neutral)  weak gauge bosons. For low-energy (hadronic) processes, the heavy gauge bosons decouple from the theory leading to 
effective parity-violating four-fermion interactions. The effective interactions 
resulting from the exchange of charged gauge bosons induce, for example, the 
beta-decay of the muon and the neutron, while the exchange of neutral gauge 
bosons gives rise to various parity-violating four-quark operators.  

Despite this theoretical foundation, the manifestation of the $P$-violating 
four-quark operators in hadronic and nuclear systems is not fully understood. 
The problem arises mainly from the nonperturbative nature of QCD at low energies.  
In order to circumvent this problem, the   nucleon-nucleon ($N\!N$)
interaction has been parametrized in the past through $P$-violating meson exchanges with adjustable strengths. Nevertheless, theoretically allowed ranges for the coupling constants could be estimated. 
This meson-exchange model is usually called the DDH-framework 
\cite{Desplanques:1979hn}.
Given enough experimental input the unknown couplings can be determined and 
other processes can then be predicted. However, the extractions of the coupling
constants from different experiments seem to be in disagreement, although a recent study \cite{Holsteinreview} shows
that a consistent picture does emerge if only results from few-body experiments are used in the analysis (for  recent
reviews, see Refs.~\cite{Holsteinreview,Schindler:2013yua}).

 In the last three decades tremendous progress has been made in understanding 
low-energy strong interactions by the application of effective field theories
(EFTs). By writing down the most general Lagrangian for the relevant
low-energy degrees of freedom that is consistent with the symmetries of the underlying
theory, \textit{i.e} QCD, one obtains an effective field theory, called
chiral-perturbation theory ($\chi$PT), which is a low-energy equivalent of
QCD (in the sense of fulfilling the same chiral Ward identities) (for a
pedagogical review, see Ref.~\cite{Bernard:2006gx}). $\chi$PT has
a major advantage that observables can be calculated perturbatively in the form of an 
expansion in $p/\Lambda_\chi$, where $p$ is the
typical momentum of the process and $\Lambda_\chi \sim 1$ GeV the 
chiral symmetry breaking scale. In principle calculations can be performed up
to any order although in practice the number of unknown low-energy constants
(LECs) increases quickly which limits the predictive power. Another
success of  $\chi$PT is the explanation of the hierarchy and the form of
multi-nucleon interactions with respect to $N\!N$ interactions. The strong
$N\!N$ potential has been derived up to next-to-next-to-next-to-leading order
(N$^3$LO) and describes the $N\!N$ experimental database with a similar quality
as the phenomenological ``high-precision" potentials (for recent reviews, see 
Refs.~\cite{Epelbaum:2008ga,Machleidt:2011zz}).

The application of $\chi$PT has led to a derivation of the effective $P$-violating 
$N\!N$ potential. At leading order (LO) this potential consists of one-pion exchange involving as a parameter the weak pion-nucleon coupling constant $h_\pi$  \cite{Kaplan:1992vj}. At next-to-leading
order (NLO) corrections appear due to $P$-violating two-pion exchange \cite{Zhu, KaiserPodd} and five $P$-odd $N\!N$ contact interactions \cite{Zhu, Savage:1998rx,  Savage:2000iv,  Girlanda:2008ts} representing short-range dynamics (one for each $S \leftrightarrow P$ wave transition). These corrections are suppressed by two powers of 
$p/\Lambda_\chi$. Additional interactions involving external photons appear also at this order. 

The effective $P$-violating $N\!N$ potential in combination with
phenomenological $P$-conserving potentials have been applied in several 
so-called ``hybrid'' calculations.  Full EFT calculations of $P$-violating effects in proton-proton ($pp$) scattering have only been
performed within pionless EFT in which the pion is integrated out and both 
$P$-conserving and $P$-violating effects are described by $N\!N$ contact 
interactions \cite{Phillips:2008hn}. Although this is a consistent framework, the absence 
of pions implies that the EFT is only applicable in the very low
energy region $E\sim M_\pi^2/(2 m_N )\simeq
10.5$\,MeV, while a pionfull treatment can be extended up to higher energies of a few hundred MeV.
Additionally, by integrating out the pion, important information on the 
chiral-symmetry properties of the $P$-violating interactions gets lost.
For a review, see Ref.~\cite{Holstein:2009zzb}.

In this paper we apply simultaneously $P$-even and $P$-odd chiral nuclear interactions in a systematic fashion. We focus on the calculation of 
the longitudinal analyzing power in $pp$ scattering for which several experimental data points exist.  There are two special features that arise for $pp$ scattering. The first one is that the 
leading-order $N\!N$ potential which causes a ${}^3S_1
\leftrightarrow {}^3 P_1$ transition  is forbidden for 
two identical protons. It becomes thus mandatory to consider the NLO $P$-odd potential
which makes the analyzing power dependent on two independent LECs. Secondly, 
the presence of the Coulomb interaction complicates the calculation. We will 
discuss both issues in detail. Our main goal is to perform a careful extraction 
of the two LECs and compare these with theoretical estimates. 

The present paper is organized as follows. In Sec.~\ref{sec:pot} we give the
parity-violating $N\!N$ potential at NLO and summarize the present knowledge of
the weak pion-nucleon coupling $h_\pi$. In Sec.~\ref{sec:asp} we discuss
the Lippmann-Schwinger equation to solve the scattering problem in
the presence of the Coulomb interaction and define the pertinent longitudinal
analyzing power $A_z$ that measures the parity violation. Sec.~\ref{sec:res}
gives a detailed discussion of the extraction of the $P$-odd LECs from the data at low and 
intermediate energies. Sec.~\ref{sec:sum} contains a short summary and conclusions.

\section{Parity-even and parity-odd nucleon-nucleon potentials}    
\label{sec:pot}

In this paper $P$-even and $P$-odd $N\!N$ potentials as obtained in chiral effective field theory \cite{Ordonez:1993tn, Ordonez:1993tn2, Epelbaum:1998ka, Epelbaum:1998ka2}
are employed. In order to obtain a description of $N\!N$ scattering data with high precision, the chiral nucleon-nucleon potential has been extended up to N$^3$LO in Refs.~\cite{Entem:2003ft, Epelbaum:2004fk}. Both approaches differ in the regularization scheme and the treatment of the cut-off appearing in the solution of the Lippmann-Schwinger equation. An advantage of the potential of Ref.~\cite{Epelbaum:2004fk} (which is also used here) is that the cut-off can be varied over a certain range which gives a handle on theoretical uncertainties. Obviously, the N$^3$LO potential consists of many terms and we refer to
Ref.~\cite{Epelbaum:2004fk} for further details. Let us continue with presenting the $P$-violating part of the $N\!N$ potential. 

The $P$-odd $N\!N$ potential has been first derived in chiral perturbation theory in
Refs.~\cite{Kaplan:1992vj, Savage:1998rx, Savage:2000iv, Zhu}. At leading order it arises from the $P$-odd pion-nucleon interaction
\begin{equation}\label{PoddLO}
\mathcal L_{\slashPsub } = \frac{h_\pi}{\sqrt{2}} \Nb (\vec \pi\times \vec \tau)^3 N,
\end{equation}
with the coupling constant  $h_\pi$. Here $N= (p\, n)^t$ denotes the nucleon isospin-doublet, $\vec \pi$ the pion isospin-triplet, and $\vec \tau$ the isospin Pauli
matrices. In combination with the standard pseudovector parity-conserving
pion-nucleon interaction, the leading-order $P$-odd one-pion-exchange (OPE) potential follows as
\begin{equation}
V_{\text{OPE}}
= - \frac{g_{A}h_\pi}{ 2\sqrt{2} F_\pi} i(\vec \tau_1\times \vec \tau_2)^3 \frac{(\vec \sigma_1+\vec \sigma_2)\cdot \vec q }{\mpi^2+q^2},
\label{onepion}
\end{equation}
with $\vec q = \vec p - \vec p^{\,\prime}$ ($q = |\vec q\,|$), where $\vec p$ and $\vec
p^{\,\prime}$ are the relative momenta of the incoming and
outgoing nucleon pair in the center-of-mass frame. $\Fp = 92.4$\,MeV is the
pion decay constant, $\mpi = 139.57$\,MeV the charged pion mass, and $g_A=1.29$ 
the nucleon axial-vector coupling constant. By using this value of $g_A$ we
have accounted for the Goldberger-Treiman discrepancy \cite{Epelbaum:2004fk}.

It is not hard to see that this OPE potential vanishes between states of equal total
isospin and dominantly contributes to the ${}^3S_1 \leftrightarrow {}^3 P_1$
transition. The OPE potential therefore does not contribute to parity violation
in $pp$ (or $nn$) scattering. The NLO corrections to the $P$-odd potential
appear at relative order $(p/\Lambda_\chi)^2$ and consist, among other
contributions, of two-pion-exchange (TPE) diagrams \cite{Zhu, KaiserPodd}. 
The TPE contributions come in the form of triangle, box, and crossed-box
diagrams. The triangle diagrams lead to the same isospin operator $(\vec \tau_1\times \vec \tau_2)^3$ as 
the OPE potential and do therefore not contribute to $pp$ 
scattering. Apart from a contribution with the same isospin-operator, the box 
and crossed-box diagrams sum up to 
\begin{eqnarray}\label{twopion}
V_{\text{TPE}}=  \frac{\sqrt{2}g^3_A  h_\pi}{(4\pi \Fp)^2\Fp}
\left[ i (\vec \tau_1+ \vec \tau_2)^3\, (\vec \sigma_1\times \vec
  \sigma_2)\cdot \vec q \, \right] L(q, \Lambda_S)\,,
\end{eqnarray}
in terms of the loop function 
\begin{equation}\label{TPEfunc}
L(q,\Lambda_S)= \frac{\w}{2q} \ln \left(\frac{\Lambda_S^2 \w^2 +q^2 s^2 
+ 2 \Lambda_S s \w q}{4\mpi^2(\Lambda_S^2+q^2)}\right)\,,\qquad \w 
= \sqrt{4\mpi^2+q^2}, \qquad s = \sqrt{\Lambda_S^2-4\mpi^2}~.
\end{equation}
Following Ref.~\cite{Epelbaum:2004fk} we have used the method of spectral regularization 
\cite{Epelbaum:2003gr} to regularize the finite part of the pion-loop. The $P$-even $N\!N$ potential
has been regularized in the same way with a spectral cut-off $\Lambda_S$. 

The TPE diagrams are divergent and counter terms are necessary in order to absorb these divergences. 
Such counter terms naturally arise within chiral EFT and appear as $N\!N$ contact
interactions at the same order as the TPE potential \cite{Zhu}. In principle, 
five independent contact interactions appear \cite{Girlanda:2008ts} but only 
one linear combination enters in $pp$ scattering. Writing this combination 
as $C$ gives the following contribution to the $P$-odd potential
\begin{eqnarray}\label{ct}
V_{\text{CT}}=\frac{C}{2\Fp \Lambda_\chi^2}\left[i (\vec \tau_1+ 
\vec \tau_2)^3 (\vec \sigma_1\times \vec \sigma_2)\cdot \vec q\,\right],
\end{eqnarray}
where $\Lambda_\chi = 1$ GeV is the chiral symmetry breaking scale. The factor $(\Fp \Lambda_\chi^2)^{-1}$ is 
inserted in order to make $C$ dimensionless. 

At the order of the TPE diagrams and counter terms, there appear 
corrections to the one-pion exchange $V_{\rm OPE}$ proportional to the quark mass. These 
corrections can be absorbed into coupling constant $h_\pi$. In the power-counting scheme of 
Ref.~\cite{Epelbaum:2004fk}, relativistic and isospin-breaking corrections
appear at higher order in the potential. 

Summarizing, the relevant $P$-odd potential in the case of $pp$ scattering at NLO 
is simply given by $V_{\rm TPE}+V_{\rm CT}$  in Eqs.~\eqref{twopion} and \eqref{ct}. 

\subsection{Estimates and limits of {\boldmath$h_\pi$}}
\label{estimates}

In an EFT, the LECs corresponding to the various effective interactions are a priori
unknown and need to be determined by fitting them to experimental
data. In the present case the microscopic theory is well-known,
\textit{i.e.} QCD supplemented with $P$-violating four-quark operators, which
means that one can attempt to calculate the LECs directly. This is a highly
non-trivial task due to the nonperturbativeness of QCD at low energies.
Despite this difficulty, several approaches exist to tackle this problem. 

Clearly the most simple one is the use of naive-dimensional analysis (NDA)
\cite{NDA, NDA2} which gives the following estimates
\begin{equation}
h_\pi \sim C \sim \mathcal O(G_F \Fp \Lambda_\chi) \sim 10^{-6},
\end{equation} 
in terms of the Fermi coupling constant $G_F$. This should be seen as an 
order-of-magnitude estimate, providing a rough scale for the size of parity violation 
in hadronic systems.

In the original DDH paper \cite{Desplanques:1979hn}, the authors have attempted to 
estimate $h_\pi$, and several other LECs associated with heavier mesons, 
using $SU(6)$ symmetry arguments and the quark model. They have found a range of 
reasonable values for $h_\pi$: 
\begin{equation}
0\leq h_\pi \leq 1.2\cdot 10^{-6},
\end{equation}
and a ``best" value of $h_\pi \simeq 4.6 \cdot 10^{-7}$, consistent with the 
NDA estimate. 

The authors of Ref. \cite{Kaiser:1989fd} have calculated several $P$-violating 
meson-nucleon vertices in a framework of a non-linear chiral Lagrangian where the 
nucleon emerges as a soliton. They have obtained significantly 
smaller values for $h_\pi \simeq 0.2 \cdot 10^{-7}$. This approach
simultaneously predicts the strong meson-nucleon coupling constants which were 
found to be in good agreement with phenomenological boson-exchange models. 
In Ref. \cite{Meissner:1998pu}, the calculation of $h_\pi$ has been sharpened based
on a three-flavor Skyrme model calculation with the result $h_\pi \simeq 1 
\cdot 10^{-7}$, which lies in between the DDH best value and the results of 
Ref.~\cite{Kaiser:1989fd}. 

Recently, the first lattice QCD calculation has been made for $h_\pi$ using a 
lattice size of $2.5\,\mathrm{fm}$ and a pion mass $M_\pi \simeq 
389\,\mathrm{MeV}$, finding the result
\begin{equation}
h_\pi = \left(1.1\pm0.5\,(\mathrm{stat}) \pm 0.5
\,(\mathrm{sys})\right)\cdot 10^{-7},
\end{equation}
which is also rather small with respect to the DDH range~\cite{hpilatt}. It should be noted that this result does not contain contributions from disconnected diagrams nor was the 
result extrapolated to the physical pion mass. 

The smaller estimates seem to be in better agreement with data.  Experiments 
on $\gamma$-ray emission from ${}^{18} F$ set the rather strong upper 
limit \cite{Adelberger:1983zz, Adelberger:1983zz2}
\begin{equation}
h_\pi < 1.3 \cdot 10^{-7}~.
\end{equation}
Although calculations for nuclei bring
in additional uncertainties, in this case these can to a certain extent be ``cancelled out" 
by comparison with the analogous $\beta$-decay of ${}^{18}\mathrm{Ne}$   \cite{Haxton:1981sf}.

Historically, the calculation of the longitudinal asymmetry in $pp$ scattering
has not been done in terms of $h_\pi$ because, as mentioned above, the OPE 
potential does not contribute.  Within the modern EFT approach, this argument 
is no longer valid because $h_\pi$ contributes via the two-pion-exchange potential. So far, these 
contributions have been considered in a hybrid approach in Refs.~\cite{Liu:2006dm, Partanen:2012qw}.
In this paper, we investigate within a full EFT approach which ranges of
$h_\pi$ and $C$ are consistent with existing data and how these ranges relate to 
the above estimates and limits.

\section{Aspects of the calculation}
\label{sec:asp}

We apply the following form of the non-relativistic Lippmann-Schwinger (LS) equation 
in momentum space
\begin{equation}\label{LS}
T^{l^\prime l\, s^\prime s}_j (p^\prime, p, E) = V^{l^\prime l\, s^\prime s}_j
(p^\prime, p) +\sum_{l^{\prime \prime}\,s^{\prime \prime}} \int_0^\infty 
dp^{\prime \prime}\, V^{l^\prime l^{\prime \prime}\, s^\prime s^{\prime
    \prime}}_j (p^\prime, p^{\prime \prime})\left(\frac{p^{\prime \prime\,2}}
{E- p^{\prime \prime\,2}/m_p + i\epsilon}\right) T^{l^{\prime \prime} l\, 
s^{\prime \prime} s}_j (p^{\prime \prime}, p, E)~,
\end{equation}
where $E$ is the center-of-mass energy and $m_p = 938.272$~MeV is the proton mass. $T^{l^\prime l\, s^\prime s}_j$ denotes the $T$-matrix element  corresponding to conserved total angular momentum $j$ for states with initial and final 
orbital angular momentum (spin) $l$ ($s)$ and $l^\prime$ ($s^\prime)$. 
The on-shell $T$-matrix is related to the $S$-matrix via
\begin{equation}\label{Sdef}
S^{l^\prime l\, s^\prime s}_j (E) = \delta^{l^\prime l}
\delta^{s^\prime s} 
- i\pi m_p q_0 T^{l^\prime l\, s^\prime s}_j (q_0, q_0, E)~,
\end{equation}
where $q_0 = \sqrt{ m_p E}$ is the on-shell center-of-mass momentum.  
$V^{l^\prime l\, s^\prime s}_j (p^\prime, p)$ is the partial-wave-decomposed
sum of the $P$-conserving and -violating potentials.  In order to use the form
of Eq. \eqref{LS} the $P$-odd potentials in Eqs. \eqref{onepion},
\eqref{twopion}, and \eqref{ct} need to be multiplied by $1/(2\pi)^{3}$. The 
partial-wave-decomposed $P$-violating potential is given in App.~\ref{secpwd}.

Despite the regularization of the TPE diagrams, the momentum integral in the 
LS equation is divergent. Following Ref.~\cite{Epelbaum:2004fk} we regularize 
the LS equation by multiplying the potential by a regulator function 
\begin{equation}
V^{l^\prime l\, s^\prime s}_j (p^\prime, p) \to
\mathrm{exp}[-p^{\prime\,6}/\Lambda^6]\,\, V^{l^\prime l\, s^\prime s}_j
(p^\prime, p)
\,\,\mathrm{exp}[-p^{6}/\Lambda^6]~, 
\end{equation}
where $\Lambda$ is a momentum cut-off. This regulator has the advantage that 
it does not influence the partial-wave decomposition ensuring that the
potential acts in the same channels as before applying the regulator. Although 
$\Lambda$ can in principle be any high-energy scale, it seems to make little 
sense to pick $\Lambda$ larger than the chiral symmetry breaking scale 
$\Lambda_\chi\sim 1$ GeV. We vary $\Lambda$ between $450$ and $600$ MeV in
order to quantify the theoretical uncertainty of the calculation. 

Because we consider $pp$ scattering, it is necessary to include the Coulomb 
interaction. To do so, we follow the approach of Ref.~\cite{Vincent:1974zz} 
which was used in Refs.~\cite{Walzl:2000cx,Carlson:2001ma, Epelbaum:2004fk} (the treatment of
the Coulomb interaction in a pionless EFT was discussed in Ref. \cite{Kong:1999sf}).
The potential is separated into a short- and long-range part
\begin{equation}
V = V_{\mathrm{short}} + V_{\mathrm{long}}~,
\end{equation}
where $V_{\mathrm{short}}$ is the sum of the strong and weak potentials and 
$V_{\mathrm{long}}$ the Coulomb potential. At a certain range $R$, the effects 
of the short-range potential can be neglected such that for $r\geq R$
\begin{equation}
V = V_{\mathrm{long}}~.
\end{equation}
At such distances, the wave functions are simply the Coulomb asymptotic states 
expressed in terms of a linear combination of regular ($F$) and irregular 
($G$) Coulomb functions. 

For $r<R$, the total potential is given by the strong and weak potentials and 
the Fourier-transformed Coulomb potential integrated up to $r=R$
\begin{equation}\label{VC}
V_C(|p^\prime-p|) = \int_0^R d^3 r\,e^{i(\vec p^{\,\prime}-\vec p)\cdot \vec r}\,\frac{\alpha}{r}~,
\end{equation}
where $\alpha = 1/137.036$ is the fine-structure constant. The LS equation is
then solved with the potential
\begin{equation}
V = V_P + V_{\slashPsub} + V_C~,
\end{equation}
to obtain the $T_{\mathrm{short}}$- and $S_{\mathrm{short}}$-matrices. Here,
$V_P$ and $V_{\slashPsub}$ are, respectively, the $P$-conserving and 
-violating potentials. We solve the LS equation in two different ways by 
treating the $P$-odd potential both perturbatively and nonperturbatively. 
We have verified that for a small enough $P$-odd potential (as is the case in 
nature where the $P$-odd potential is smaller than the strong potential by 
approximately seven orders of magnitude) both approaches give identical
results. Technical details on the solutions are provided in App.~\ref{LSsolve}. 

At the boundary of the sphere with radius $R$ the two solutions with
potentials $V_{\mathrm{short}}+V_{\mathrm{long}}$ and $V_{\mathrm{long}}$ need
to match. This can be done by demanding the logarithmic derivative of both 
solutions to be equal. The actual matching is most conveniently done via 
the $K$-matrix which is related to the $S$-matrix by
\begin{equation}\label{Kdef}
K_{\mathrm{short}} = -\frac{i}{m_p q_0}(1-S_{\mathrm{short}})(1+S_{\mathrm{short}})^{-1},
\end{equation}
where $K_{\mathrm{short}}$ and $S_{\mathrm{short}}$ are $4\times 4$ matrices 
analogous to Eq.~\eqref{Tmat}. The $K$-matrix $K_{\mathrm{long}}$ is then
obtained from the relation
\begin{eqnarray}
K_{\mathrm{long}}& =& \frac{1}{m_p q_0} \left[F(F_0 -m_p q_0
  K_{\mathrm{short}} G_0)^{-1}(F^\prime_0 -m_p q_0 K_{\mathrm{short}} 
G^\prime_0) -F^\prime \right]\nonumber\\
&& \times\left[G(F_0 -m_p q_0 K_{\mathrm{short}} G_0)^{-1}(F^\prime_0 
-m_p q_0 K_{\mathrm{short}} G^\prime_0)-G^\prime \right]^{-1},
\end{eqnarray}
where we introduced  $4\times4 $ matrices containing the (ir)regular Coulomb functions
\begin{equation}
F(r) = \bma 
F_{j-1}(r)&0&0&0\\
0&F_{j+1}(r)&0&0\\
0&0&F_{j}(r)&0\\
0&0&0&F_j(r)\ema,
\,
G(r) = \bma 
G_{j-1}(r)&0&0&0\\
0&G_{j+1}(r)&0&0\\
0&0&G_{j}(r)&0\\
0&0&0&G_j(r)\ema~.
\end{equation}
Here, $F_0$ and $G_0$ denote the Coulomb functions in the presence of zero
charge and the ${}^\prime$ implies differentiation with respect to $r$. We 
still need to specify at what range $R$ we perform the matching. It cannot be 
too low, since the short-range potential needs to vanish but too large radii 
give problems due to rapid oscillations induced by Eq.~\eqref{VC}. Here we 
follow Ref.~\cite{Epelbaum:2004fk} and perform the matching at $R =12$~fm.

Once $K_{\mathrm{long}}$ has been determined, the $S_{\mathrm{long}}$- and 
$T_{\mathrm{long}}$-matrices in the presence of the Coulomb interaction 
with respect to the Coulomb asymptotic states can be obtained via the 
inverse relations of Eqs.~\eqref{Sdef} and \eqref{Kdef}. In what follows
below, we always refer to these quantities and drop the subscript.

\subsection{Scattering amplitude}
The solution of the $T$-matrix can be used to calculate the scattering 
amplitude $M^{m_1^\prime m_2^\prime\,m_1 m_2}$ where $m_1$ ($m_2$) and 
$m_1^\prime$ ($m_2^\prime$) are the third component of the spin of the
incoming and outgoing protons. To do so, we first write the on-shell
$T$-matrix in a different basis
\begin{eqnarray}
T^{s^\prime m_s^\prime\, s ms}(\hat p^\prime,\,\hat p, E) &=& \sum_{j m_j
  l^\prime l} T^{l^\prime l\, s^\prime s}_j(E) Y_{l^\prime\, m_j-m_s^\prime}
(\hat p^\prime) Y^\star_{l^\prime\, m_j-m_s}(\hat p)\nonumber\\
&&\times C(l^\prime\,s^\prime\,j;m_j-m_s^\prime\,m_s^\prime\,m_j)
C(l\,s\,j;m_j-m_s\,m_s\,m_j)~ ,
\end{eqnarray}
in terms of the spherical harmonics $Y_{lm}(\Omega)$ and the Clebsch-Gordan 
coefficients $C(j_1\,j_2\,j;m_1\,m_2\,m_1+m_2)$. In the results below, unless stated otherwise, we perform the sum over the total angular momentum up to $j\leq4$. Contributions from higher values of $j$ are negligible. The choice $\hat p = \hat z$ implies
\[ Y^\star_{l^\prime\, m_j-m_s}(\hat p) = \sqrt{\frac{2l+1}{4\pi}}\delta^{m_j m_s},\]
such that
\begin{eqnarray}
T^{s^\prime m_s^\prime\, s ms}(\theta, E) &=& \sum_{j l^\prime l}  
\sqrt{\frac{2l+1}{4\pi}} T^{l^\prime l\, s^\prime s}_j (E)  
Y_{l^\prime\, m_s-m_s^\prime}(\theta)\nonumber\\
&&\times  C(l^\prime\,s^\prime\,j;m_s-m_s^\prime\,m_s^\prime\,m_s)C(l\,s\,j;0\,m_s\,m_s) ,
\end{eqnarray}
where $\theta$ is the scattering angle in the center-of-mass frame. 
A final basis change then gives 
\begin{equation} 
T^{m_1^\prime m_2^\prime\,m_1 m_2}(\theta, E) = \sum_{s^\prime s\,m_s^\prime
  m_s} C(\frac{1}{2}\,\frac{1}{2}\,s;m_1\,m_2\,m_s)C(\frac{1}{2}\,
\frac{1}{2}\,s^\prime;m_1^\prime\,m_2^\prime\,m_s^\prime)\,
T^{s^\prime m_s^\prime\, s ms}(\theta, E).
\end{equation}

For identical particles, the amplitude $M$ is related to the on-shell $T$-matrix via
\begin{equation} \label{Mstrong}
 M^{m_1^\prime m_2^\prime\,m_1 m_2}(\theta, E) = -m_p (4\pi^2) \,
T^{m_1^\prime m_2^\prime\,m_1 m_2}(\theta, E) \times\frac{1}{2}[1+ 
(-1)^{l+s}]\times\frac{1}{2}[1+(-1)^{l^\prime+s^\prime}],
\end{equation}
where the factors between square brackets are there to ensure the Pauli principle. 

So far, we have calculated the scattering amplitude in the presence of the
Coulomb interaction with respect to the Coulomb asymptotic states. Due to 
screening effects, experiments are performed with free asymptotic states  
and, in order to compare with the experimental data, this discrepancy 
needs to be remedied. We follow the approach outlined in Ref.~\cite{Taylor}. 
First, the amplitude obtains a Coulomb phase factor
\begin{equation}\label{MstrongC}
 M^{m_1^\prime m_2^\prime\,m_1 m_2}(\theta, E)  \to M^{m_1^\prime
   m_2^\prime\,m_1 m_2}(\theta, E) e^{i\sigma_l} e^{i\sigma_{l^\prime}}~,
\end{equation}
in terms of $\sigma_l = \mathrm{arg}[\Gamma(l+1+i\eta)]$ and $\eta = 
\alpha m_p/(2q_0)$. Second, we add the anti-symmetrized Coulomb amplitude 
\begin{eqnarray}
M_C^{m_1^\prime m_2^\prime\,m_1 m_2}(\theta, E)&=& \sum_{s m_s}
C(\frac{1}{2}\,\frac{1}{2}\,s;m_1\,m_2\,m_s)C(\frac{1}{2}\,\frac{1}{2}\,
s;m_1^\prime\,m_2^\prime\,m_s)\nonumber\\
&&\times \left( f_C(\theta,E) + (-1)^s f_C(\pi-\theta,E) \right),
\end{eqnarray}
where
\begin{equation}
 f_C(E,\theta) = -\frac{\eta}{2q_0 \sin^2{\theta/2}}e^{i\left(2\sigma_0
-\eta \ln \sin^2 \theta/2 \right)}~.
\end{equation}
The total amplitude $\bar M$ thus becomes
\begin{equation}
\bar M^{m_1^\prime m_2^\prime\,m_1 m_2}(\theta, E) = 
M^{m_1^\prime m_2^\prime\,m_1 m_2}(\theta, E) e^{i\sigma_l}
e^{i\sigma_{l^\prime}} 
+ M_C^{m_1^\prime m_2^\prime\,m_1 m_2}(\theta, E)~.
\end{equation}

Before continuing, it is instructive to look at the total Coulomb cross 
section given by
\begin{eqnarray}\label{Ccross}
\sigma_C(E) &=& \int d\Omega \frac{1}{4}\mathrm{tr}\left[ M_C(\theta, E) 
M_C^\dagger(\theta, E)\right]\nonumber\\
&=& \frac{\pi\eta^2}{ q_0^2}\left(\frac{1}{\sin^2 \theta_c/2}
-\frac{1}{\cos^2 \theta_c/2}+\frac{1}{\eta}\sin\left[2 \eta\ln\left\{\tan 
\theta_c/2\right\}\right]\right)~.
\end{eqnarray}
Here, we introduced a small critical opening angle $\theta_c$ in order to keep 
the result finite. For small values of $\theta_c$ and/or $E$ the Coulomb cross 
section becomes very large which has important consequences for the
longitudinal asymmetry, to which we now turn. 
 
\subsection{Longitudinal analyzing power}

The longitudinal asymmetry is defined as the difference in cross section
between the scattering of an unpolarized target with a beam with positive and 
negative helicity normalized to the sum of both cross sections. 
Mathematically this becomes
\begin{equation}\label{AP}
 A_z(\theta, E)= \frac{\mathrm{tr}\left[\bar M(\theta, E)\,\sigma_z\, \bar
     M(\theta, E)^\dagger\right]}{\mathrm{tr}\left[\bar M(\theta, E)\!\bar
     M(\theta, E)^\dagger\right]}~,
\end{equation}
where $\sigma_z = \sigma_z^{(1)}\otimes \,\mathcal I_2^{(2)}$ is the Kronecker
product of the third Pauli matrix and the two-dimensional unit matrix,
corresponding to a longitudinally polarized beam and an unpolarized target. 
Experiments typically measure over a certain angular range and report the 
integrated asymmetry
\begin{equation}\label{AP2}
\bar A_z (E)= \frac{\int_{\theta_1}^{\theta_2}d\!\cos\theta\,\mathrm{tr}
\left[\bar M(\theta, E)\,\sigma_z\, \bar M(\theta, E)^\dagger\right]}
{\int_{\theta_1}^{\theta_2}d\!\cos\theta\,\mathrm{tr}\left[\bar M(\theta, E)\!
\bar M(\theta, E)^\dagger\right]}.
\end{equation}

Transmission experiments, on the other hand, measure the transmitted beam from 
which the total cross section (apart from scattering under angles smaller than 
some critical angle $\theta_c$) is inferred \cite{Driscoll:1988hg}. That is, in the absence of inelastic scattering, 
they report
\begin{equation}\label{APtrans}
\bar A_z (E)= \frac{\int_{\theta_c}^{\pi/2}d\!\cos\theta\,\mathrm{tr}
\left[\bar M(\theta, E)\,\sigma_z\, \bar M(\theta, E)^\dagger\right]}
{\int_{\theta_c}^{\pi/2}d\!\cos\theta\,\mathrm{tr}\left[\bar M(\theta, E)\!
\bar M(\theta, E)^\dagger\right]}.
\end{equation}

As can be seen from Eq.~\eqref{Ccross} for small values of $\theta_c$, the
Coulomb cross section becomes very large which suppresses the integrated 
longitudinal asymmetry. The results can therefore become very sensitive to 
the specific value of $\theta_c$ \cite{Driscoll:1988hg}. We will discuss 
this in more detail below. 

\section{Comparison with experiments}
\label{sec:res}

The longitudinal asymmetry in $pp$ scattering has been measured at several
energies. The experiments with highest precision are the Bonn experiment at 
$13.6$~MeV \cite{Eversheim:1991tg, Eversheim:1991tg2}, the PSI experiment at $45$~MeV 
\cite{Kistryn:1987tq}, and the TRIUMF experiment at $221$ MeV
\cite{Berdoz:2001nu} (all energies are lab energies). The first two
experiments are scattering experiments which report $\bar A_z$ over an angular 
range of, respectively, $20^{\circ}$-$78^{\circ}$ and $23^{\circ}$-$52^{\circ}$ (lab coordinates)
\begin{eqnarray}\label{exp12}
\bar A_z (13.6\,\mathrm{MeV}) &=& (-0.93\pm 0.21)\cdot 10^{-7}~,\\
\bar A_z (45\,\mathrm{MeV}) &=& (-1.50\pm 0.22)\cdot  10^{-7}~.
\end{eqnarray}
The results of the calculations at $13.6$~MeV are almost independent of the 
angular range as long as a small forward angles are excluded. We have
confirmed that using the range $23^{\circ}$-$52^{\circ}$ gives results within 
$3\%$ of using the actual range measured in the experiment. For presentation 
purposes in most plots below we use the range of the $45$~MeV experiment.

The experiment at $221$~MeV is a transmission experiment and reports
\begin{eqnarray}\label{exp3}
\bar A_z (221\,\mathrm{MeV}) &=& (0.84\pm0.34)\cdot 10^{-7}~.
\end{eqnarray}

\begin{figure}[t]
\centering
\includegraphics[scale=1.0]{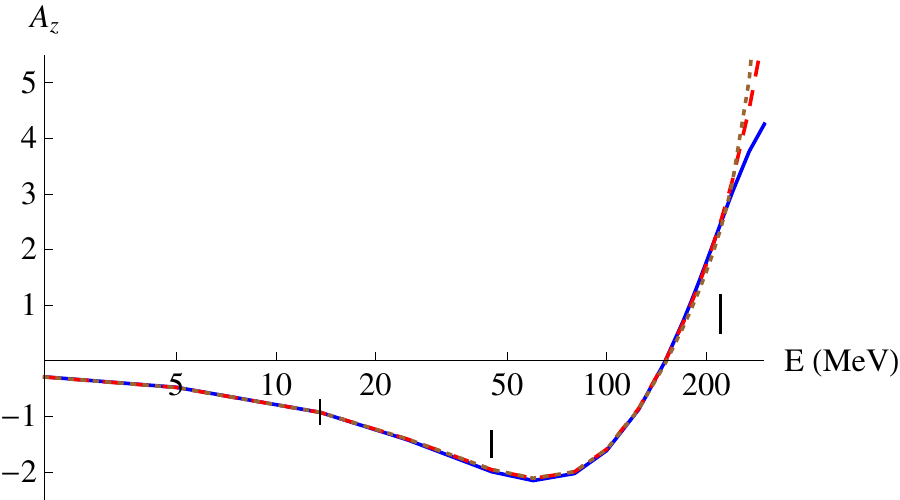}
\caption{The integrated (angular range $23^{\circ}$ to $52^{\circ}$ (lab)) 
asymmetry $\bar A_z$ (in units of $10^{-7}$) as a function of laboratory 
energy. The energy in MeV is plotted on a logarithmic scale. The blue 
(solid), red (dashed), and brown (dotted) lines correspond to the different 
cut-off combinations in Eq.~\eqref{cut}. The DDH ``best'' value, 
$h_\pi = 4.6\cdot 10^{-7}$, has been used.}
 \label{DDHhpi}
\end{figure} 

\subsection{Fit of the counter term}
\begin{figure}[t]
  \centering
 \includegraphics[width=0.49\textwidth]{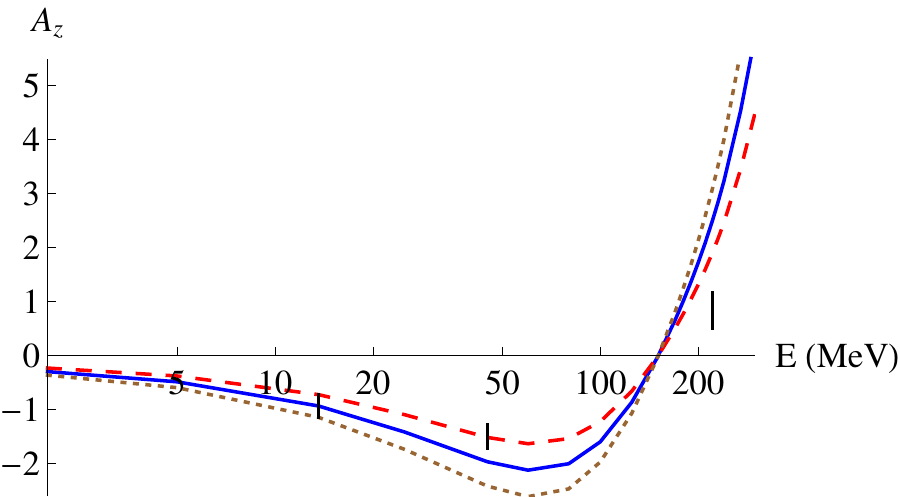}   
 \includegraphics[width=0.49\textwidth]{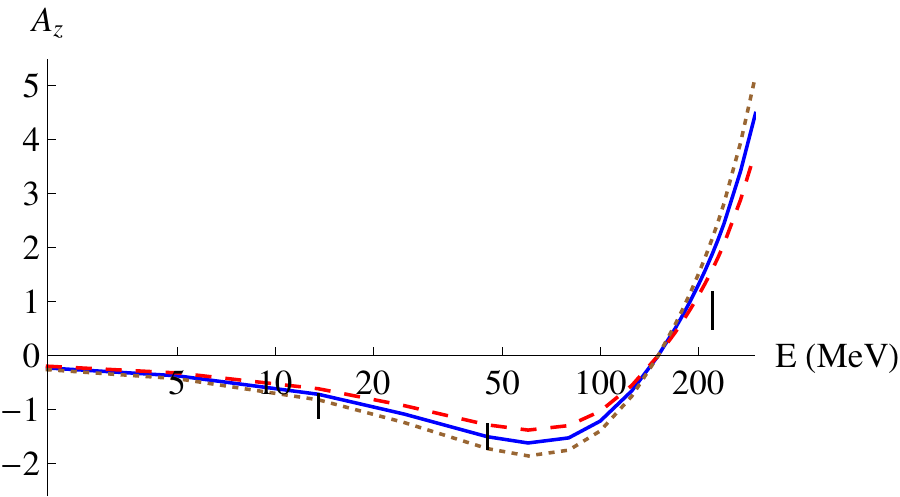}
\caption{The integrated (angular range $23^{\circ}$ to $52^{\circ}$ (lab)) 
asymmetry $\bar A_z$ (in units of $10^{-7}$) as a function of lab energy. 
The blue (solid), red (dashed), and brown (dotted) lines correspond to a 
fit to the central value of the first (left plot) or second (right plot) 
data point, the central value plus one standard deviation, and the central 
value minus one standard deviation. The intermediate cut-off combination in 
Eq.~\eqref{cut} has been used and $h_\pi = 4.6\cdot 10^{-7}$.}
 \label{DDHhpi2}
\end{figure}

The calculation of $\bar A_z(E)$ depends on two unknown LECs: the 
pion-nucleon coupling constant $h_\pi$ and the nucleon-nucleon coupling 
constant $C$. We require two data points in order to fit both LECs. Before
doing so, we first study the results if we use what is known as the 
DDH ``best'' value $h_\pi = 4.6\cdot 10^{-7}$. We fit the LEC $C$ to the 
central value of the lowest-energy data point. In order to probe the 
cut-off dependence we perform the fit for three different cut-off combinations 
(all values in MeV)
\begin{equation}\label{cut}
\{\Lambda,\,\Lambda_S\}=\{450,\,500\},\,\{550,\,600\},\,\{600,\,700\}~,
\end{equation}
to obtain the three fits
\begin{equation}
C = \{-4.5,\,-5.1,\,-5.5\}\cdot 10^{-6}~,
\end{equation}
corresponding to 
cut-off dependence of 
approximately $10\%$. Using the DDH value for $h_\pi$ and the fit values for 
$C$, the prediction for $\bar A_z(E)$, integrated from $23^{\circ}$ to
$52^{\circ}$, is shown in Fig.~\ref{DDHhpi}. First of all, the cut-off 
dependence of $\bar A_z$ over the whole relevant energy range is very small, 
only becoming visible at energies above $221$~MeV. Second, the predictions 
seem to disagree significantly with the $221$~MeV data point. This, however, 
is of no concern. The reason being that the measurement at $221$~MeV
corresponds to a different angular range which, as we will discuss below, 
has important consequences.  Finally, the predictions somewhat overestimate
$|\bar A_z(45\,\mathrm{MeV})|$.  To study this in more detail, we now take the 
intermediate cut-off combination and fit $C$ to the central value plus or
minus one standard deviation of the first data point. We obtain the following fits 
\begin{equation}\label{fit1}
C = \{-4.3,\,-5.1,\,-6.0\}\cdot 10^{-6}~.
\end{equation}
The predictions are shown in the 
left panel of Fig.~\ref{DDHhpi2}. 
The second data point is now well described within the experimental uncertainty. 
Alternatively, we can fit $C$ to the data point at $45$~MeV. Doing so with the 
intermediate cut-off combination gives the fit 
\begin{equation}
C = \{-4.1,\,-4.5,\,-4.9\}\cdot 10^{-6}~,
\end{equation}
and the predictions in the right panel of Fig.~\ref{DDHhpi2}. Again both 
low-energy data points are well described within the experimental uncertainty. 

\begin{figure}[t]
\centering
 \includegraphics[width=0.49\textwidth]{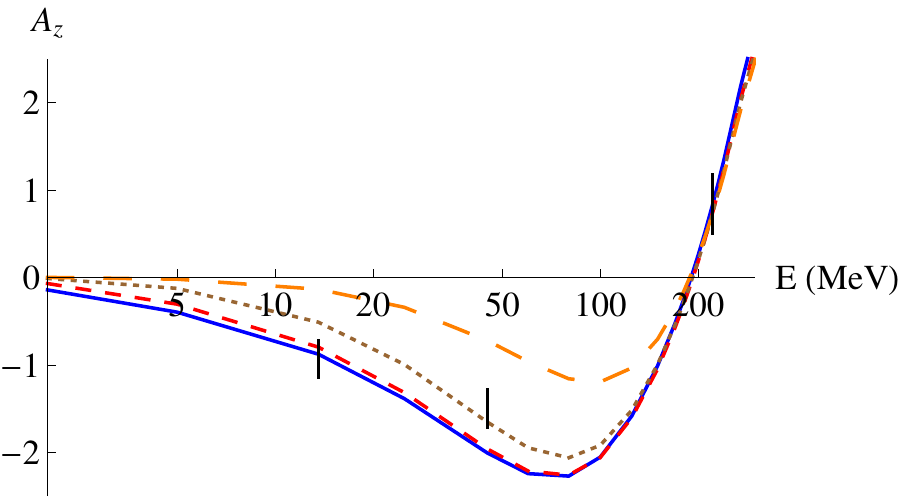}   
 \includegraphics[width=0.49\textwidth]{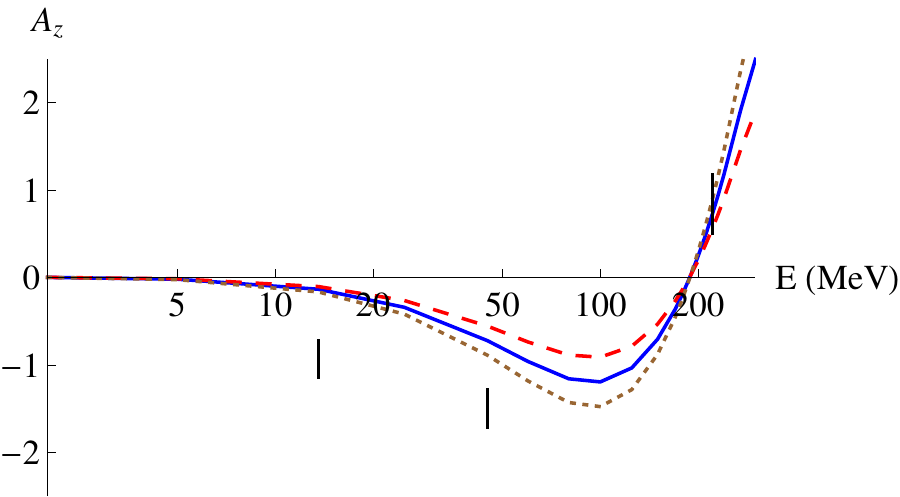}
\caption{The integrated (angular range $\theta_c$ to $90^{\circ}$ (CM frame)) 
asymmetry $\bar A_z$ (in units of $10^{-7}$) as a function of lab-energy. 
Left plot: the blue (solid), red (dashed), brown (dotted), and orange 
(long-dashed) lines correspond, respectively, to $\theta_c = 15^\circ,
\,10^\circ,\,5^\circ,\,2^\circ$. 
Right plot: The integrated asymmetry with $\theta_c=2^\circ$. The blue 
(solid), red (dashed), and brown (dotted) lines correspond to a fit through 
the first data point plus or minus one standard deviation. In both plots 
the intermediate cut-off combination has been used and $h_\pi = 4.6\cdot 10^{-7}$.}
 \label{thetacritDDH}
\end{figure} 

In order to include the third data point into the analysis we need to
integrate over a different angular range. The experiment at $221$~MeV 
measures almost the whole cross section apart from scattering under angles 
smaller than a small critical angle $\theta_c$. In the left graph of 
Fig.~\ref{thetacritDDH}, we plot $\bar A_z(E)$ for various values of
$\theta_c$. We use the intermediate cut-off combination, the DDH best value 
for $h_\pi$, and $C = -5.1\cdot 10^{-6}$ corresponding to a fit to the central 
value of the first data point. The graphs tells us that at low energies 
a transmission experiment would be very dependent on the critical angle, 
but at $221$~MeV there is only a small difference when varying $\theta_c$ 
between $15^\circ$ and $2^\circ$. These conclusions are in line with the observations made 
in Refs.~\cite{Driscoll:1988hg, Carlson:2001ma, Partanen:2012qw} where the 
critical angle behaviour was also studied, albeit for different $P$-even and 
-odd potentials. With the current fit parameters we predict an asymmetry as 
measured in the $221$~MeV experiment of 
\begin{equation} 
6.7 \cdot 10^{-8} \leq \bar A_z(221\,\mathrm{MeV}) \leq 7.7\cdot 10^{-8}~, 
\end{equation}
in excellent agreement with data. Here the variance, much smaller than the 
experimental uncertainty, is due to the different choices for $\theta_c$. 

It should be noted that for $\theta_c\geq 10^\circ$, the results for $\bar
A_z(13.6\,\mathrm{MeV})$ and $\bar A_z(45\,\mathrm{MeV})$ are largely
insensitive to the angular range as can be seen by comparison of 
Figs.~\ref{DDHhpi} and \ref{thetacritDDH}. At higher energies, varying 
the angular range has more impact.

In the right panel of  Fig.~\ref{thetacritDDH}, we show  $\bar A_z(E)$ 
integrated from $2^\circ$ to $90^\circ$, using the fit values in 
Eq.~\eqref{fit1} for $C$. In order to see how well this fit describes the 
three data points, one needs to look at this graph for the high-energy 
data point and the left panel of Fig.~\ref{DDHhpi2} for the two low-energy 
points. The annoyance of having to look at two plots can be avoided by 
choosing an angular range which corresponds reasonably well to all three data 
points. As discussed above, the value of $\bar A_z(E)$ at $13.6$ and $45$ MeV 
is rather insensitive to the actual angular range as long as the opening angle 
is larger than $10^\circ$, while $\bar A_z(221\,\mathrm{MeV})$ corresponds 
very well to the range $10^\circ$ to $90^\circ$. Using this range we find
indeed a good fit to all three data points. 

Although the DDH ``best'' value $h_\pi = 4.6\cdot 10^{-7}$, accompanied by one
four-nucleon operator with the LEC $C\simeq -5.1\cdot 10^{-6}$, describes the 
existing data satisfactory, this does not imply that these values correspond 
to the values taken by nature. Taking the lattice-QCD predicted value 
\cite{hpilatt}, which agrees with a Skyrme-based prediction 
\cite{Meissner:1998pu}, $h_\pi= 1\cdot 10^{-7}$ gives a fit (intermediate cut-off)
\begin{equation}\label{fitskyrme}
C = \{-3.2,\,-4.1,\,-4.9\}\cdot 10^{-6}~,
\end{equation}
and predicts asymmetries at $45$~MeV and $221$~MeV of
\begin{equation}
\label{Azsmallhpi}
\bar A_z(45\,\mathrm{MeV}) = -(1.9\pm0.45)\cdot 10^{-7},\qquad 
\bar A_z(221\,\mathrm{MeV}) = +(0.72\pm0.16)\cdot 10^{-7}~,
\end{equation} 
in agreement with the data, despite a somewhat large prediction of 
$|\bar A_z(45\,\mathrm{MeV}|$. At these small values of $h_\pi$, $\bar A_z(E)$ 
depends dominantly on the counter-term contributions while the TPE
contributions are smaller by an order of magnitude.

Finally, rather large values of $h_\pi$ are allowed as well. Using $h_\pi = 
1.5\cdot 10^{-6}$, which lies somewhat above the DDH reasonable range,  
gives the following fit for $C$ (intermediate cut-off)
\begin{equation}\label{fitlarge}
C = \{-7.4,\,-8.3,\,-9.2\}\cdot 10^{-6}~,
\end{equation}
and
\begin{equation}\label{Azlargehpi}
\bar A_z(45\,\mathrm{MeV}) = -(1.5\pm0.45)\cdot 10^{-7}~,\qquad 
\bar A_z(221\,\mathrm{MeV}) = +(0.54\pm0.17)\cdot 10^{-7}~,
\end{equation} 
again consistent with the data. It should be noted that with these 
large values for $h_\pi$ the cut-off dependence of the results for 
$\bar A_z(E)$ becomes significant (approximately $20\%$ at $45$~MeV and $50\%$
at $221$~MeV). This uncertainty is not captured in the error margins of 
Eq.~\eqref{Azlargehpi}. The increase of the cut-off dependence is due to the 
larger value of $h_\pi$. The counter term only absorbs cut-off dependence in
the lowest partial-wave transition ${}^1S_0\leftrightarrow {}^3 P_0$ while, 
at higher energies, the TPE potential also contributes to transitions with 
larger total angular momentum.

What might be surprising is that in the results of Eqs.~\eqref{Azsmallhpi} 
and \eqref{Azlargehpi},  the uncertainty of $\bar A_z(221\,\mathrm{MeV})$ is 
smaller than that of $\bar A_z(45\,\mathrm{MeV})$. We will come back to this peculiar behaviour later.

\subsection{Fit of both low-energy constants}
So far we have been inspired by theoretical estimates of the pion-nucleon 
coupling constant $h_\pi$. However, we have also seen that a relatively large 
range of values for $h_\pi$ describes the data properly, assuming the counter 
term is fitted to one of the data points. In this section we assume no, 
a priori, knowledge of $h_\pi$ and fit both LECs to the data points. 
We first fit the LECs to the low-energy data points and predict the third. The
reason for fitting first to the low-energy points is that at these energies 
we can expect our EFT analysis to be most accurate while at higher energies 
higher-order corrections might start playing a role. 

  \begin{figure}[t]
  \centering
 \includegraphics[width=0.49\textwidth]{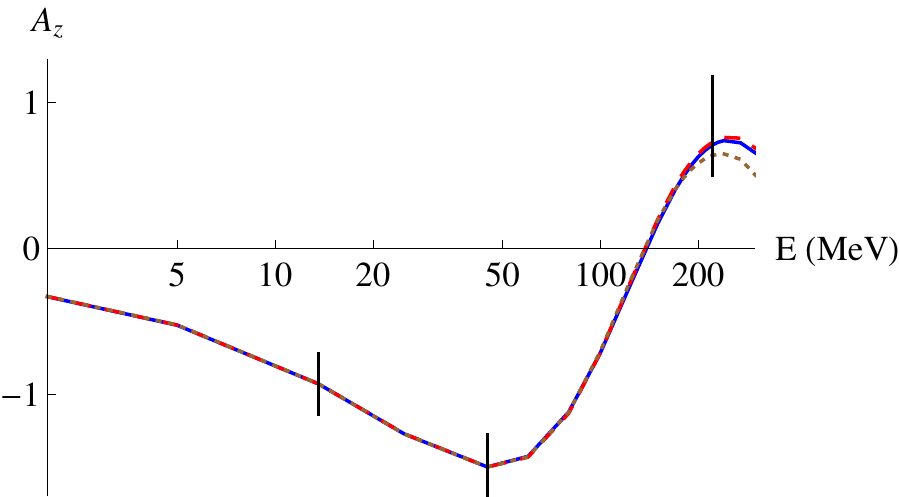}   
 \includegraphics[width=0.49\textwidth]{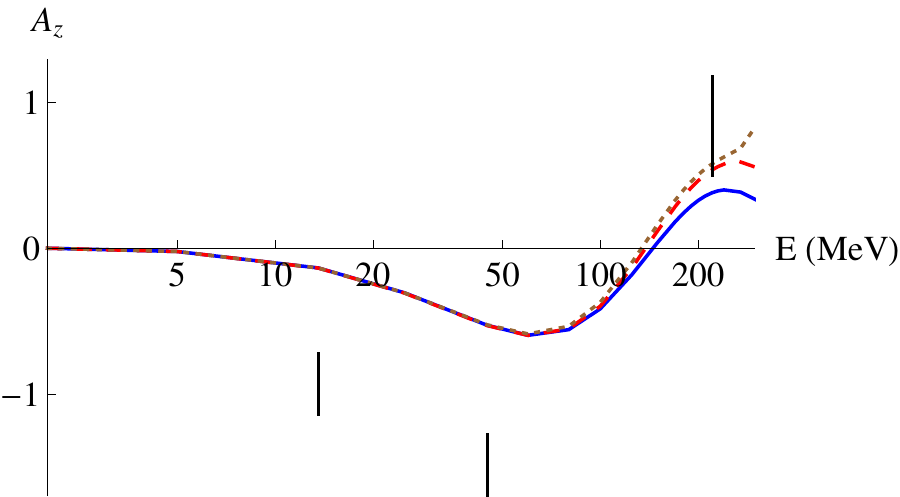} 
 \caption{The integrated asymmetry $\bar A_z$ (in units of $10^{-7}$) as a
   function of the lab energy (left plots: angular range $23^{\circ}$ to
   $52^{\circ}$ (lab), right plots: angular range $2^{\circ}$ to $90^{\circ}$
   (CM)). The blue (solid), red (dashed), and brown (dotted) lines correspond 
   to the three cut-off combinations in Eq.~\eqref{cut}. $h_\pi$ and $C$ have
   been fitted to the central values of the low-energy data points.}
 \label{Fitboth}
\end{figure}

Fitting $h_\pi$ and $C$ to the central value of the first two data points, 
while using the three cut-off combinations in Eq.~\eqref{cut}, gives the following fits
\begin{eqnarray}
h_\pi &=& \{1.3,\,1.5,\,2.0\}\cdot 10^{-6},\nonumber\\
C &=& \{-7.5,\,-8.3,\,-10\}\cdot 10^{-6}.
\end{eqnarray}
The fit of $h_\pi$ is remarkably large with respect to the estimated values
and in stark disagreement with the experimental limits given in
Sec.~\ref{estimates}. Before investigating this in more detail, we show the
plot of the asymmetry in Fig.~\ref{Fitboth} for the relevant angular ranges. 
First of all, we note that the cut-off dependence has increased with respect 
to the results in Fig.~\ref{DDHhpi} which is due to the increase of $h_\pi$. 
The cut-off dependence is still much smaller than the experimental
uncertainty. Secondly, the prediction for the high-energy data point is on the 
low side but the theoretical and experimental error bands do overlap. 

  \begin{figure}[t]
  \centering
 \includegraphics[width=0.49\textwidth]{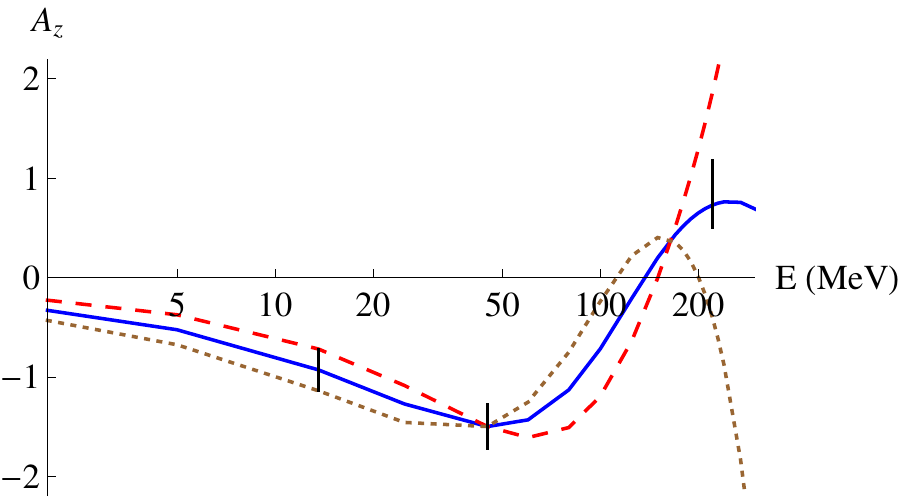}   
 \includegraphics[width=0.49\textwidth]{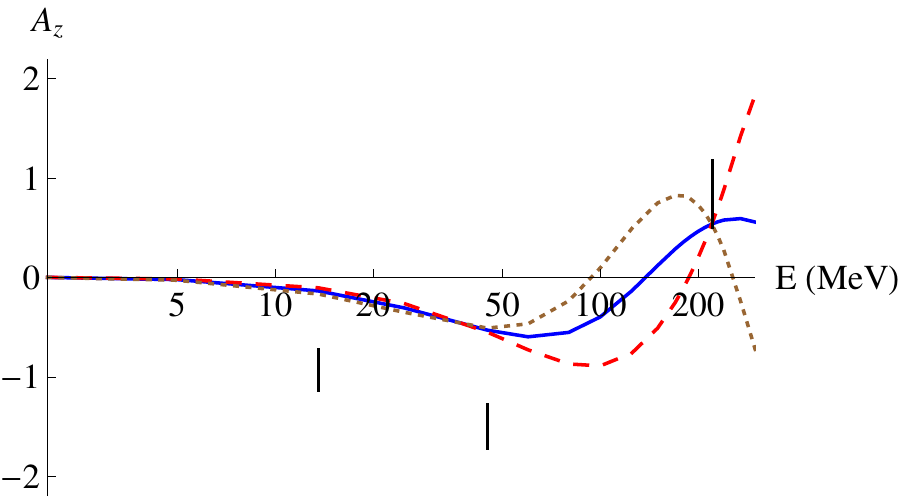} 
 \caption{The integrated asymmetry $\bar A_z$ (in units of $10^{-7}$) as a 
  function of the lab energy (left plots: angular range $23^{\circ}$ to
  $52^{\circ}$ (lab), right plots: angular range $2^{\circ}$ to $90^{\circ}$ 
  (CM)). The blue (solid), red (dashed), and brown (dotted) lines correspond 
   to the three different fits. For details, see text.}
 \label{Fitboth2}
\end{figure}

In Fig.~\ref{Fitboth2} we show similar graphs, but we now fitted the LECs to 
the central value plus or minus one standard deviation of the first data point 
and the central value of the second data point. The intermediate cut-off 
combination has been used. The range of the LECs becomes very large
\begin{eqnarray}\label{fit2}
h_\pi &=& \{0.14,\,1.5,\,2.8\}\cdot 10^{-6}~,\nonumber\\
C &=& \{-3.3,\,-8.3,\,-13\}\cdot 10^{-6}~,
\end{eqnarray}
spanning more than an order of magnitude. The smallest value of $h_\pi$ 
is not far from the experimental limit and rather close to the smaller 
estimates in Sec.~\ref{estimates}. Noteworthy is that, despite the huge
variance in coupling constants, all three fits almost exactly cross at the 
energy of the third data point. We will come back to this in detail later.

If we simultaneously vary the second data point by plus or minus one 
standard deviation and the cut-off combination we obtain the following allowed values
\begin{eqnarray}\label{fit2points}
h_\pi &=& (1.7\pm 2.5 )\cdot 10^{-6}~,\nonumber\\
C &=& (-9.3\pm 10)\cdot 10^{-6}~.
\end{eqnarray}
Here we only give an estimate for the allowed range, in Sec.~\ref{fit3points} 
we perform a more detailed analysis. The fits of $h_\pi$ and $C$ are, of
course, highly correlated which can be seen from the contours in
Fig.~\ref{chi}. The fits tell us that small values of $h_\pi \sim 10^{-7}$ 
are not ruled out, however they are definitely not favored. Most fits prefer 
an $h_\pi \sim 10^{-6}$ which is of the order of the NDA estimate but, 
as mentioned, such values disagree strongly with the experimental upper
limits. The cut-off dependence of the fits is modest.

\subsection{Crossing points}\label{sec:cross}

\begin{figure}[!t]
\centering
\includegraphics[width=0.49\textwidth]{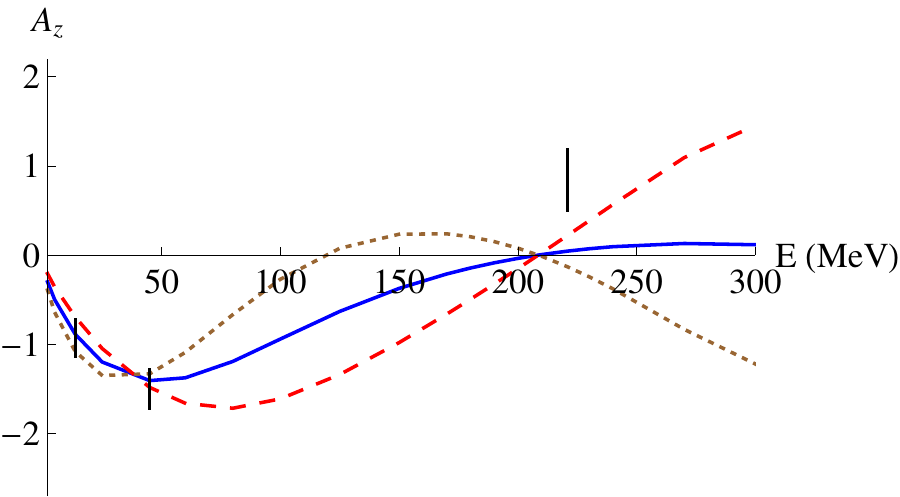} 
\includegraphics[width=0.49\textwidth]{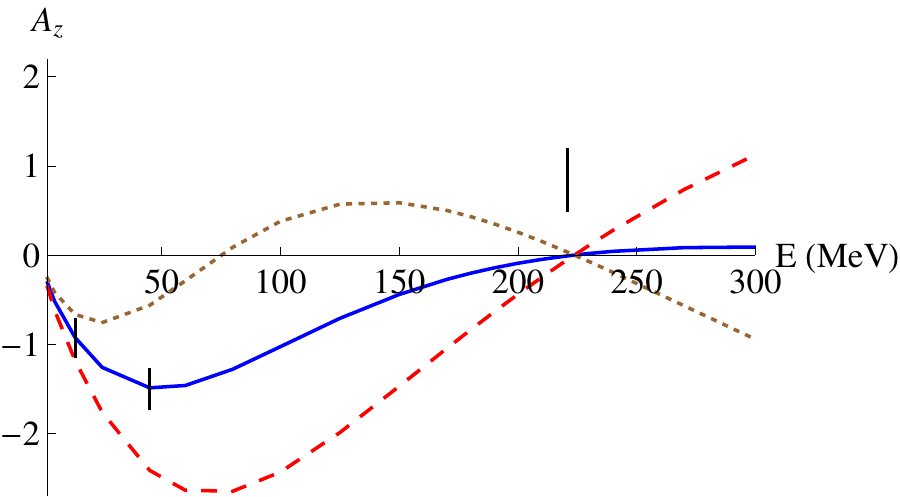}  
\includegraphics[width=0.49\textwidth]{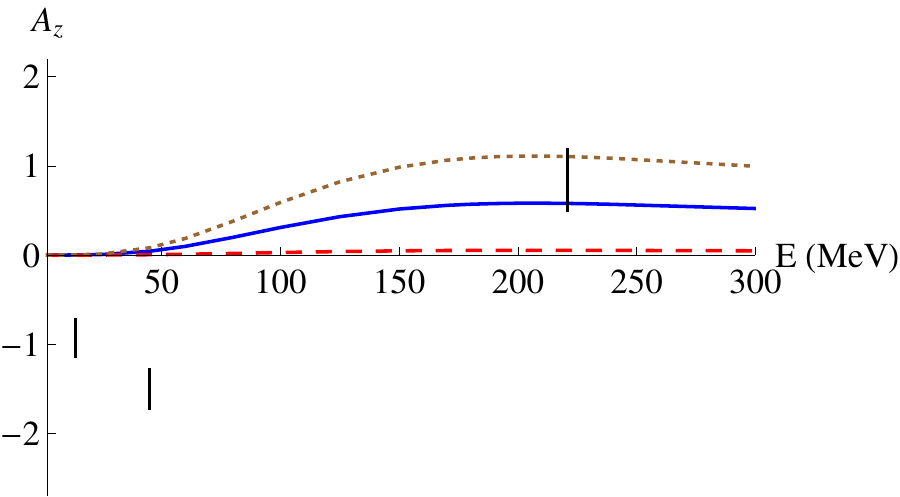} 
\includegraphics[width=0.49\textwidth]{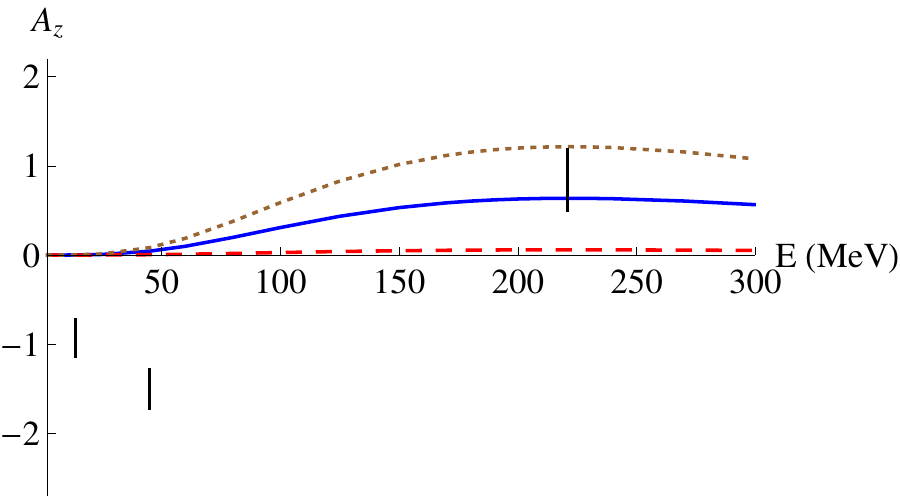} 
\includegraphics[width=0.49\textwidth]{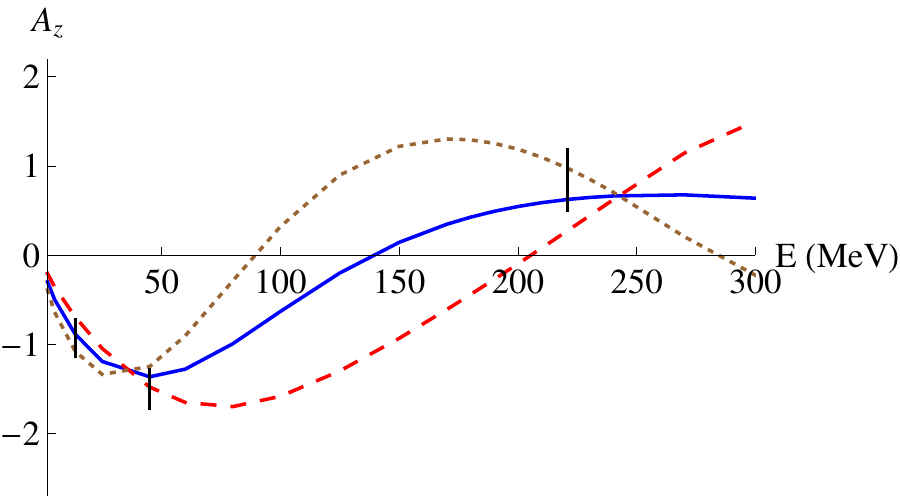}  
\includegraphics[width=0.49\textwidth]{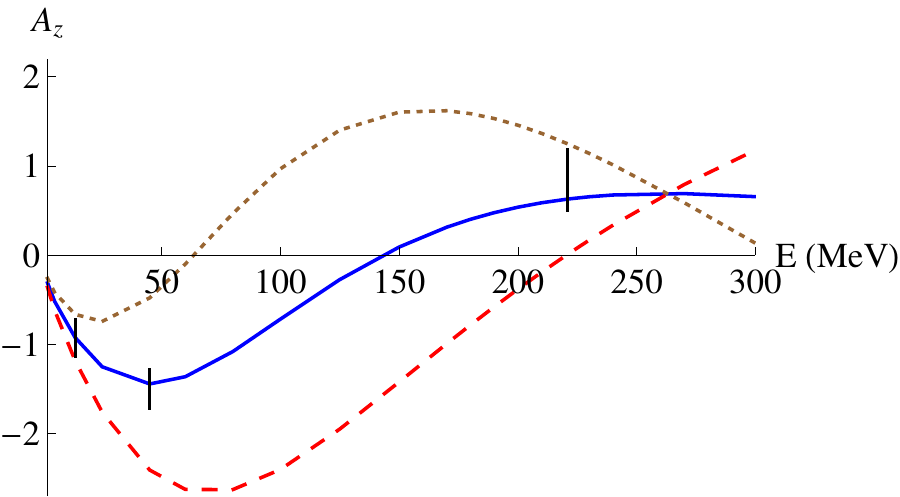}
\caption{The integrated (angular range $0^{\circ}$ to $90^{\circ}$ 
(CM frame)) asymmetry $\bar A_z$ (in units of $10^{-7}$) as a function 
of the lab energy. The Coulomb amplitude $M_C$ is neglected. In the top-plot 
only $j=0$ $P$-odd transitions are taken into account, in the middle plot all 
$P$-odd transitions for $0<j\leq4$, and in the bottom all $P$-odd transition with $j\leq 4$. The plots on the left correspond to the N$^3$LO potential with 
intermediate cut-off, and the plots on the right to the NijmII potential.}
 \label{noC}
\end{figure}

The observation that the three different fits cross in one point in the right 
plot of Fig.~\ref{Fitboth2} around $220$~MeV is somewhat surprising. In order 
to understand this behaviour it is useful to 
dissect the results in terms of different partial-wave contributions. 
For simplicity we first do the analysis without including the Coulomb
amplitude $M_C$. In Fig.~\ref{noC} we use the three fit-values in
Eq.~\eqref{fit2} for the LECs and plot the total asymmetry in the case we 
neglect $M_C$. The plot at the top-left shows the contribution coming 
from ${}^1S_0\leftrightarrow {}^3P_0$ transitions only, the middle-left plot 
shows the contribution from all $P$-odd transitions with $0<j\leq 4$, and the 
bottom-left plot shows the complete asymmetry and is, therefore, the sum 
of the two plots above. 

The top-left plot shows that the $j=0$ contributions vanish at an energy of 
approximately $210$~MeV. This well-known behaviour 
\cite{Driscoll:1988hg, Driscoll:1989jv, Carlson:2001ma} is due to the
vanishing of $\delta_{{}^1S_0} + \delta_{{}^3P_0}$ (where $\delta$ denotes the 
strong phase shifts) at this particular energy. In fact, this zero-crossing
was one of the main reasons for the chosen energy of the TRIUMF experiment.  
It should be noted that the exact point of crossing can vary by $\pm5$~MeV 
for the different cut-off combinations. Also, more phenomenological potentials 
such as the NijmII potential \cite{Stoks:1994wp} have the zero-crossing around $225$~MeV (see the 
top-right plot). This dependence on the details of the strong potential
already indicates a larger theoretical uncertainty.

The second observation is that between $200$ and $270$~MeV the $j=0$ 
contributions depends almost linearly on the energy. Around these energies 
the asymmetry is proportional to
\begin{equation}  
\sin (\delta_{{}^1S_0} - \delta_{{}^3P_0}) \sin ( \delta_{{}^1S_0} + 
\delta_{{}^3P_0} +\sigma_0 +\sigma_1) \simeq  (\delta_{{}^1S_0} 
- \delta_{{}^3P_0})(\delta_{{}^1S_0} + \delta_{{}^3P_0})~,
\end{equation}
which indeed shows a linear behaviour from $190$~MeV onwards. Here, we 
neglected $\sigma_0+\sigma_1\simeq -0.1^\circ$ ($\sigma_l$ is defined right below Eq. \eqref{MstrongC}) which is much smaller than the individual 
strong phase shifts. The linearity is not affected significantly by the 
energy dependence of the $P$-odd potentials or total cross section which 
are fairly constant in this range. Since the $j=0$ contributions depend 
both on $h_\pi$ and $C$, the contributions can be parametrized by
\begin{equation}
\bar A_z(E)^i_{j=0} = (a\,h_\pi^i + b\,C^i)(E-E_0)~, 
\end{equation}
where the index $i$ specifies which fit parameters are used, $E_0$ is the 
energy of the zero-crossing point, and $a$ and $b$ are fit-independent 
constants which can be determined from the slopes of the lines in the top 
plot. This parametrization only holds in the range where our assumptions 
regarding the strong phase-shift behaviour hold, which is more-or-less between 
$200$ and $270$~MeV. 

The middle-left plot of Fig.~\ref{noC} shows that the contributions from the
higher partial waves (which are to good approximation dominated by the 
${}^3P_2\leftrightarrow {}^1D_2$ transitions \cite{Carlson:2001ma}) are 
almost constant between $200$ and $270$~MeV, due to the fact that the 
$j=2$ strong phase-shifts and mixing angle hardly vary over this range. 
Since the $j>0$ contributions depend only on $h_\pi$ the total asymmetry 
can be parametrized as
\begin{equation}
\bar A_z(E)^i= (a\,h_\pi^i + b\,C^i)(E-E_0) + c\,h_\pi^i~, 
\end{equation}
introducing one more constant $c$ which can be obtained from the height 
of the lines in the middle plot. 

In order to have a crossing point as seen in the bottom-left plot, the 
following equation should hold for any two fits $i$ and $j$ 
\begin{equation}\label{crosseq}
0 = \bar A_z(E_0^\prime)^i - \bar A_z(E_0^\prime)^j 
= [a\,(h_\pi^i-h_\pi^j) + b\,(C^i-C^j)](E_0^\prime-E_0) + c\,(h_\pi^i-h_\pi^j), 
\end{equation}
at a certain energy $E_0^\prime$. In general, such an equation does not hold 
for all $i$ and $j$. However, due to the fact that $P$ violation is a 
perturbative effect, the fitting procedure will always provide a linear 
relation between the two LECs, as can be clearly seen from the contours 
in Fig.~\ref{chi}. Using $C^i = \alpha h_\pi^i + \beta$ in
Eq.~\eqref{crosseq}, in terms of two new constants,  gives 
\begin{equation}\label{crosseq2}
0 =  [a + b\alpha](E_0^\prime-E_0) + c~.
\end{equation}
This relation needs to hold, within the energy range where the approximations 
are valid, in order for a crossing point to exist. The constants $a$, $b$, and 
$c$ can be determined from the graphs giving $a \simeq -3.5 \cdot 10^{-3}\,
\mathrm{MeV}^{-1}$, $b \simeq -6.5 \cdot 10^{-4}\,\mathrm{MeV}^{-1}$, and $c  
\simeq 3.0\cdot 10^{-2}$, while $\alpha \simeq -3.5$ can be obtained from 
Fig.~\ref{chi}. For these values, we obtain
\begin{equation}
E_0^\prime-E_0 \simeq 25\,\mathrm{MeV}~,
\end{equation}
which implies a crossing point at approximately $E_0 +25 \,\mathrm{MeV} \simeq 
235\,\mathrm{MeV}$, close to the actual crossing point and within the range
where the approximations hold.  

The above analysis shows that the existence of crossing point mostly hinges on 
the energy-dependence of the relevant strong phase shifts. As such, the
existence of these points is quite insensitive to the strong potential used 
as long as it roughly predicts the correct energy scaling of the phase
shifts. The actual location of the crossing point, on the other hand, is much 
more sensitive to details of the potential, in particular to the exact point 
where $\delta_{{}^1S_0} + \delta_{{}^3P_0} =0$, but also on the precise sizes
of the phase shifts. To illustrate this, we show the same graphs as before 
but now using the NijmII potential (note that, for illustrative purposes, 
we use the same values for $h_\pi$ and $C$ and did not refit them), on the 
right-hand side of Fig.~\ref{noC}. The crossing point still exists, but now 
appears around $265$~MeV and is shifted by $40$~MeV from $E_0 \simeq 225\,
\mathrm{MeV}$. 

\begin{figure}[!t]
\centering
\includegraphics[width=0.49\textwidth]{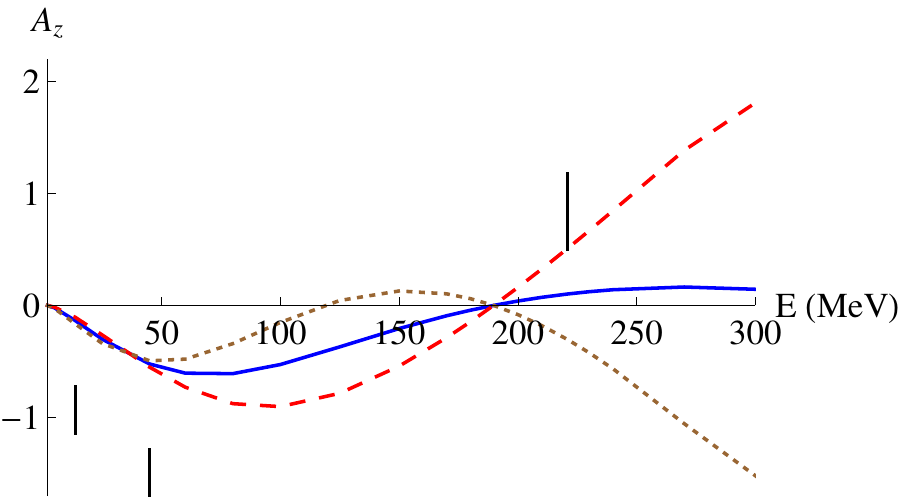}  
\includegraphics[width=0.49\textwidth]{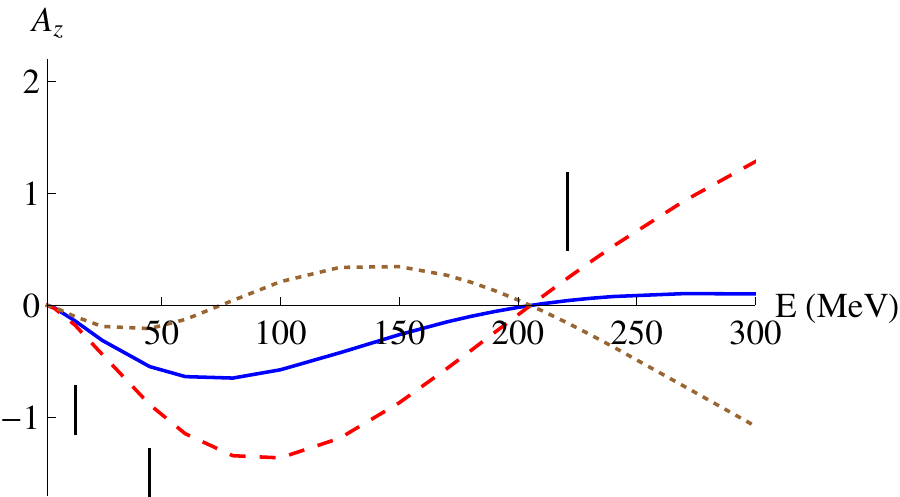} 
\includegraphics[width=0.49\textwidth]{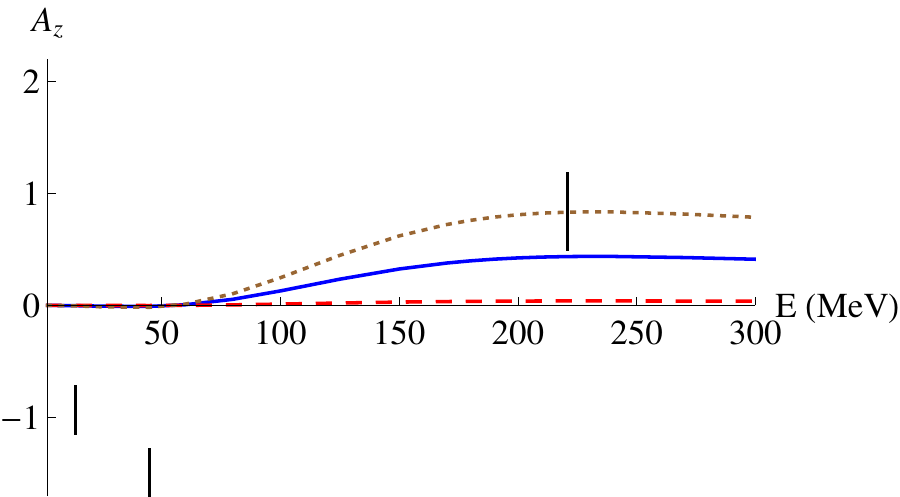}  
\includegraphics[width=0.49\textwidth]{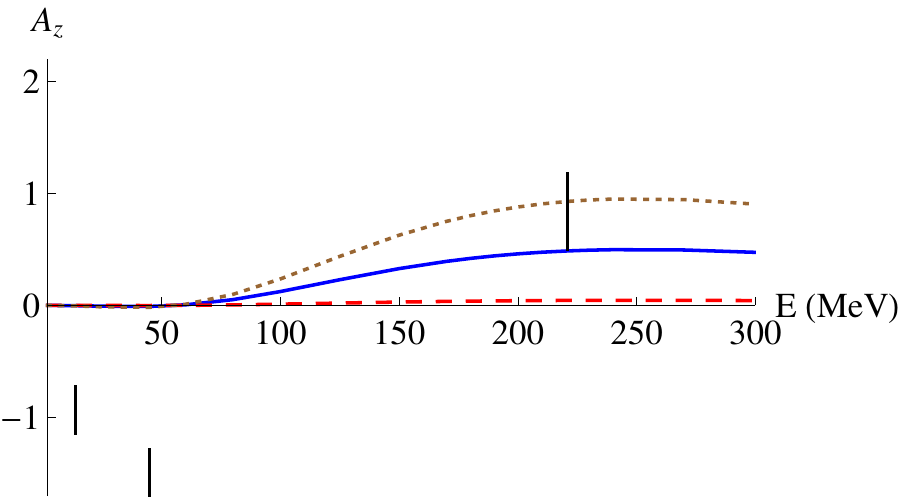} 
\includegraphics[width=0.49\textwidth]{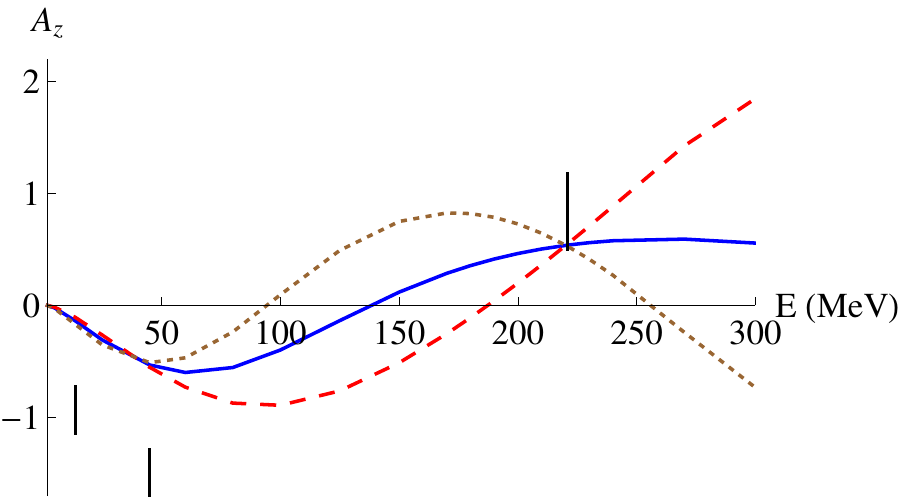} 
\includegraphics[width=0.49\textwidth]{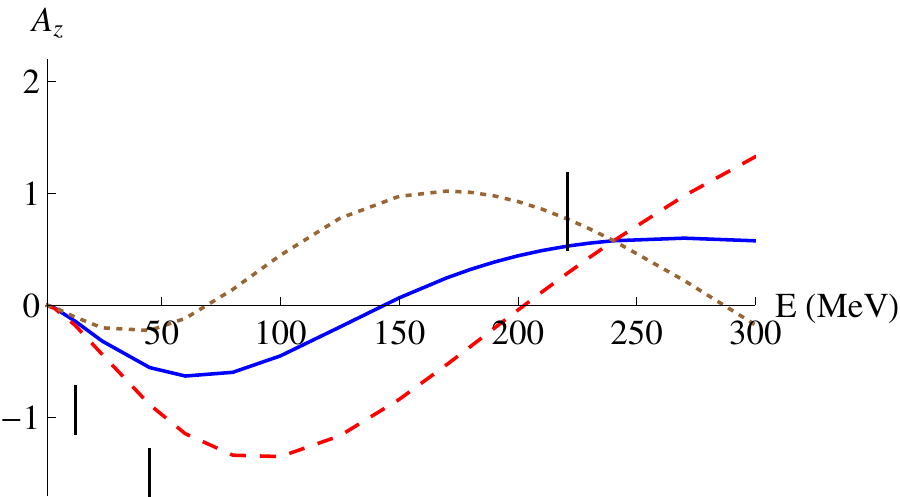}
\caption{The integrated (angular range $2^{\circ}$ to $90^{\circ}$ 
(CM frame)) asymmetry $\bar A_z$ (in units of $10^{-7}$) as a function of 
the lab energy. The Coulomb amplitude $M_C$ is included. In the top-plot 
only $j=0$ $P$-odd transitions are taken into account, in the middle plot all 
$P$-odd transitions for $0<j\leq4$, and in the bottom all $P$-odd transition with $j\leq 4$. The plots on the left correspond to the N$^3$LO potential with 
intermediate cut-off, and the plots on the right to the NijmII potential.  }
 \label{withC}
\end{figure} 

The analysis so far has neglected the Coulomb amplitude. In Fig.~\ref{withC}, 
we show the same plots (using the same fits) which do take $M_C$ into
account. We have used $\theta_c =2 ^\circ$ as the critical angle in order to
avoid the Coulomb divergence. The plots are very similar to the ones in
Fig.~\ref{noC}. The main difference is the location of the  zero-crossing
points in the plots at the top, and the crossing points in the plots at the bottom. 
All these points are shifted to lower energies by approximately $20$~MeV. 
As shown in Refs.~\cite{Driscoll:1988hg, Carlson:2001ma}, introducing the 
Coulomb amplitude causes the $j=0$ transitions to become proportional to
\begin{equation}  
\sin (\delta_{{}^1S_0} - \delta_{{}^3P_0}) \sin ( \delta_{{}^1S_0} 
+ \delta_{{}^3P_0}+\sigma_0 +\sigma_1 + \phi)~,
\end{equation}
where 
\begin{equation}
\phi = 2\left[ \eta\,\ln (\sin \frac{\theta_c}{2})
-\sigma_0\right]\simeq -4^\circ,
\end{equation}
around $200$~MeV lab energy and using $\theta_c = 2^\circ$. Due to this additional phase, 
the zero-crossing for $j=0$ contributions is shifted to the energy where
$\delta_{{}^1S_0} + \delta_{{}^3P_0} =4^\circ,$ which happens at an energy 
approximately $20$~MeV lower than the original zero-crossing point at $E_0$. 

Although the prediction of the total asymmetry is not influenced by a large 
amount (at least for energies larger than $100$~MeV) by the Coulomb amplitude, as can be seen by comparing the bottom plots 
in Figs.~\ref{noC} and \ref{withC}, the interpretation of the $221$~MeV data 
point in terms of partial-wave transitions has become murkier. This is best 
illustrated by looking at the right panels which correspond to the NijmII 
potential. In the plot without the Coulomb amplitude, the asymmetry is only 
due to $j>0$ transitions and thus depends only on
$h_\pi$. This was the 
reason why the experiment was done at this energy in the first place. The
Coulomb amplitude, however, shifts the zero-crossing of the $j=0$ transitions 
to $205$~MeV which means that the asymmetry at $221$~MeV obtains contributions 
from $j=0$ and $j>0$ transitions and depends on both $h_\pi$ and $C$.  
The argument that the $221$~MeV is only sensitive to $j>0$ transitions is thus 
not completely correct once the Coulomb amplitude is included, even if one 
uses phenomenological potentials which have a phase-shift cancellation at this 
energy. Of course, the same analysis holds for the chiral potential, but in 
this case the asymmetry at $221$~MeV already depends on both $j=0$ and $j>0$ 
transitions before including the Coulomb amplitude. 

The fact that, once the Coulomb amplitude has been included,  the crossing
point for the chiral potential lies almost exactly at the energy of the third 
data point, should be seen as a coincidence. Nevertheless, the observation
that, in general, the crossing point lies very close to the third data point 
implies that this point has less discriminating power with respect to the fit 
parameters than might be expected. Furthermore, the sensitivity to details of 
the strong interaction potential combined with the knowledge that the chiral 
potentials are not very accurate at these energies, means that this data point 
is hard to analyze in our EFT framework. 

\subsection{Fit through all data points}
\label{fit3points}

\begin{figure}[h]
\centering
\includegraphics[width=0.49\textwidth]{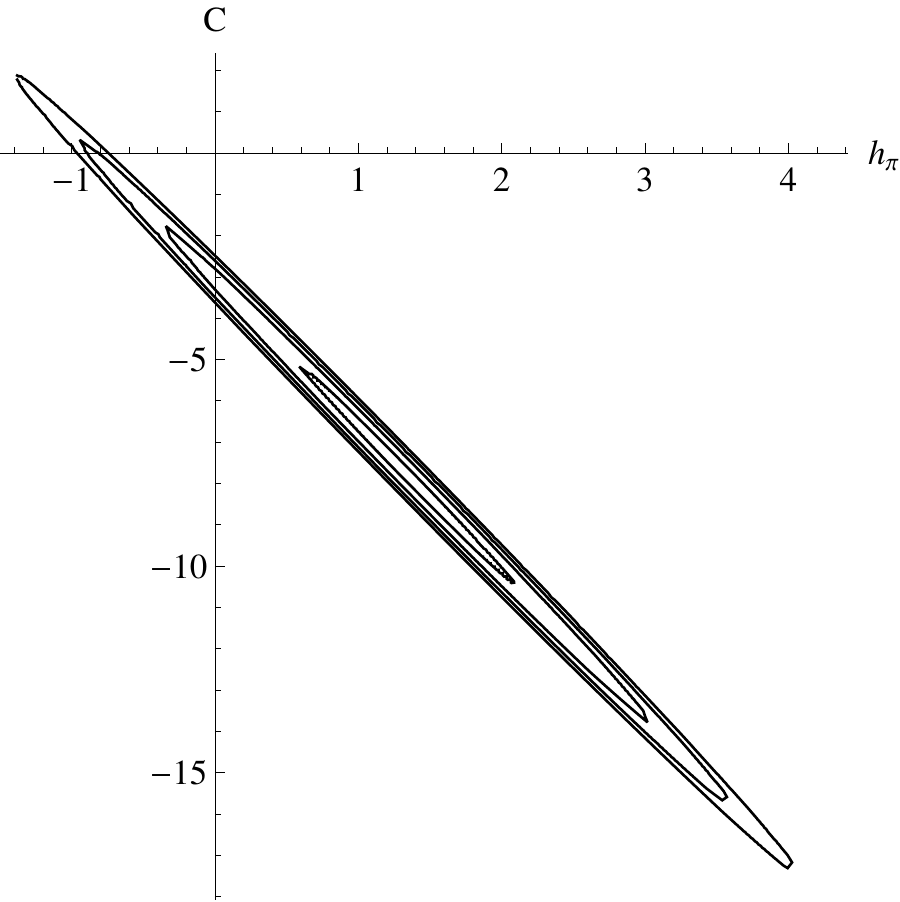}  
\includegraphics[width=0.49\textwidth]{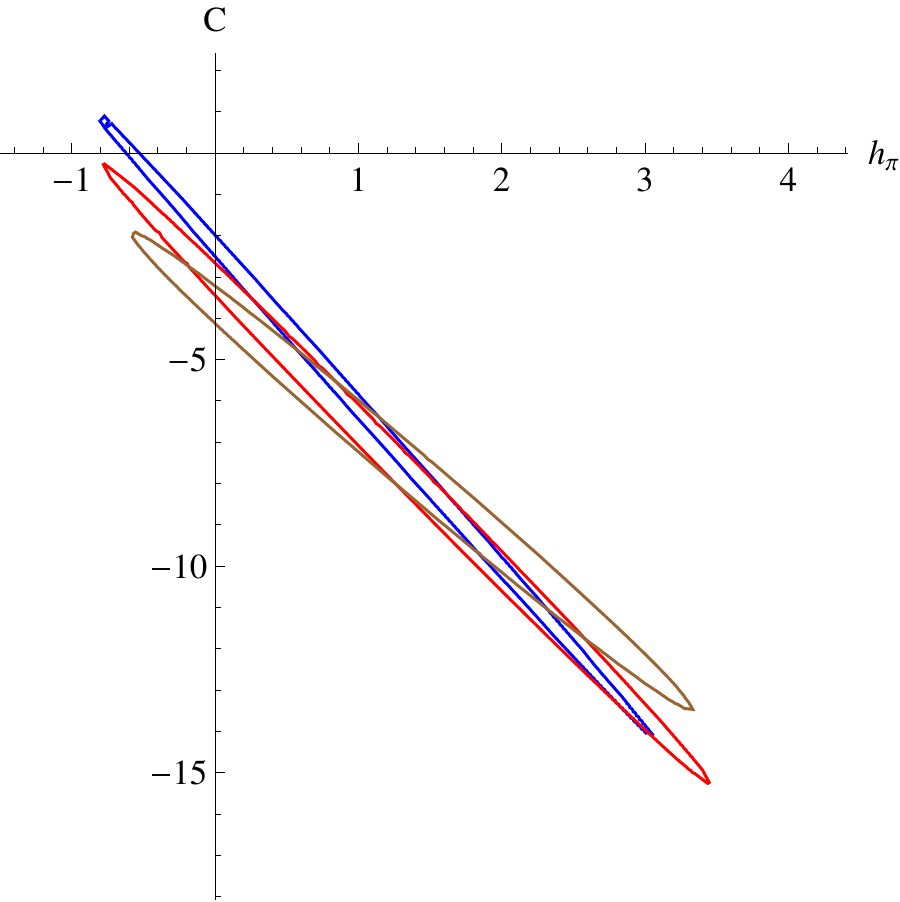}
\caption{Contours of constant $\chi^2$ in the $h_\pi-C$ plane (both in units 
of $10^{-6}$). The left plot shows contours of total $\chi^2=1,\,2,\,3,\,4$ for 
the intermediate cut-off combination, while the right plot shows contours 
of total $\chi^2=2.71$ for the three different cut-off combinations.  }
 \label{chi}
\end{figure} 
Despite the issues raised in the previous section related to the data point 
at $221$~MeV, it is still interesting to investigate a fit through all
points. In the left part of Fig.~\ref{chi}, we plot contours of constant total 
$\chi^2 = 1,\,2,\,3,\,4$ using the intermediate cut-off combination. In the
right part we study the cut-off dependence of the fit by plotting  
contours of constant total $\chi^2 =2.71$ for the three different cut-off combinations. 

From the right plot it becomes clear that the cut-off dependence of the fit is 
small since the contours mostly overlap. Second, the left plot shows that at 
the level of total $\chi^2=1$ the contour does not include small values of 
$h_\pi \sim 10^{-7}$ which are favored by theory 
\cite{Kaiser:1989fd, Meissner:1998pu, hpilatt} and the experimental data 
on ${}^{18} F$  $\gamma$-ray emission \cite{Adelberger:1983zz, Adelberger:1983zz2}. However, these values
are already included at the level of total $\chi^2=2$ and we conclude that our
analysis of the longitudinal asymmetry is consistent with 
such small values of $h_\pi$. Clearly, our analysis allows for much larger 
(and smaller) values of $h_\pi$ as well and more data is needed to further 
pinpoint the size of this important LEC. All in all, the allowed range for the 
LECs, at the total $\chi^2 =2.71$ level, is approximately
 \begin{eqnarray}\label{fit3}
h_\pi &=& (1.1\pm 2 )\cdot 10^{-6}~,\nonumber\\
C &=& (-6.5\pm 8)\cdot 10^{-6}~.
\end{eqnarray}
Although the uncertainties of the fit are reduced compared to 
Eq.~\eqref{fit2points}, the reduction is smaller than might be expected due to the 
existence of the crossing points.

With the latter comments in mind, it becomes interesting to study at which 
energies a new experiment would have most impact. A smaller energy than 
$221$~MeV is preferred because at lower energies the chiral potentials are 
more reliable. Simultaneously, the energy should be significantly higher than 
$45$~MeV in order not to overlap with the PSI experiment. An experiment at a lab energy between $100$ and $150$~MeV seems to be best suited. 
These energies have the major advantage over $221$~MeV that they are 
sufficiently far from the crossing points.

Apart from the energy, the angular range is also of importance 
\cite{Driscoll:1988hg, Carlson:2001ma}. By looking at Fig.~\ref{Fitboth2}, 
we see that a larger angular range has more discriminating power. On the 
other hand, the opening angle needs to be big enough such that there is 
no large sensitivity to small variations in $\theta_c$. It seems an experiment 
measuring from $10^\circ$ to $80^\circ$ (lab coordinates) combines the best 
of both worlds.

\section{Discussion and conclusions}
\label{sec:sum}
Historically, parity violation in hadronic processes has mostly been discussed in the one-boson-exchange framework of DDH \cite{Desplanques:1979hn}. 
In this framework, parity violation arises due to the single exchange of a
pion, $\rho$- or $\omega$-meson. In the chiral EFT approach we adopt
here, the exchange of the heavy mesons are captured by four-nucleon
contact interactions. One-pion exchange appears in both the DDH and the chiral EFT framework, however, in the latter, at the same order as the contact interactions, there are contributions due to two-pion-exchange diagrams \cite{Zhu, KaiserPodd}. Due to its isospin properties one-pion exchange  vanishes in $pp$ scattering. In the DDH framework the longitudinal asymmetry does therefore not depend on the weak pion-nucleon coupling constant $h_\pi$. Consequently, this important LEC has been often neglected in calculations of the longitudinal asymmetry. A proper low-energy description of hadronic parity violation contains two-pion-exchange diagrams which do contribute to the asymmetry in $pp$ scattering. These contributions have so far been investigated in a hybrid approach in Refs.~\cite{Liu:2006dm, Partanen:2012qw}, but the authors of these references did not extract the value of $h_\pi$. 

In this work, we reinvestigated the asymmetry in $pp$ scattering in chiral effective field theory. For the $P$-conserving $N\!N$ potential we used the N$^3$LO potential obtained from chiral effective field theory \cite{Epelbaum:2004fk} and, within the same power-counting scheme, the $P$-violating potential up to NLO. Both potentials are systematically regularized and theoretical errors due to cut-off dependence have been investigated and found to be negligible at low energies. At higher energies the uncertainty grows but is still much smaller than experimental errors. 

We have found that the $P$-odd NLO potential, consisting of TPE contributions and one four-nucleon contact term, successfully describes the existing data. The two unknown LECs can be fitted to the data at $13.6$ and $45$ MeV, and the third data point at $221$ MeV can be predicted. Unfortunately, our analysis has shown that, due to the particular energy-dependence of the strong phase shifts and the $P$-odd potential, different fits for the unknown LECs predict more-or-less similar asymmetries around $221$ MeV. This behaviour limits the discriminating power of the $221$ MeV data point and forces us to adopt a rather large allowed range for $h_\pi$ and $C$. 

The allowed range for $h_\pi = (1.1\pm 2 )\cdot 10^{-6}$ is consistent with the experimental limits obtained from $\gamma$-ray emission of ${}^{18}F$ and with theoretical estimations. However, it is clear that more experimental data is needed to reduce the uncertainty on $h_\pi$. Our analysis shows that an additional measurement of the asymmetry in the energy range of $100$ to $150$ MeV would be beneficial. This energy has some advantages over the $221$ MeV data point. The most important ones being that the chiral potentials (both the $P$-conserving and $P$-violating) are more accurate at lower energies and that such energies are sufficiently far away from the crossing points discussed in Sec.~\ref{sec:cross}. 

Additional input can, of course, come from other observables than the
$pp$ longitudinal asymmetry. In particular, the angular asymmetry in
$\vec n p\rightarrow d\gamma$ is a very promising observable although,
so far, there only exists an experimental upper bound. A major
advantage of this observable is that, in contrast with the $pp$
asymmetry, it does depend on the LO $P$-odd potential and thus cleanly
probes $h_\pi$  \cite{Kaplan:1998xi}. A caveat is that, if $h_\pi$ is really as small as
suggested, this observable might also obtain important contributions
from higher-order corrections in the form of parity-violating
four-nucleon contact or nucleon-pion-photon interactions. We plan to investigate this observable in our chiral-EFT approach in future work. 

The allowed range for $C = (-6.5\pm 8)\cdot 10^{-6}$ is harder to compare with existing literature in which these contributions are usually described via $\rho$- and $\omega$-meson exchange. The $pp$ asymmetry is typically expressed in terms of two independent combinations (one for the ${}^1S_0-{}^3P_0$ transition and one for the ${}^3P_2-{}^1D_2$ transition) of DDH couplings \cite{Holsteinreview}. By application of resonance-saturation methods these approaches can be compared, but one must be careful to not double count the TPE contributions (for details, see Refs.~\cite{Epelbaum:2001fm, Berengut:2013nh}). Here we refrain from a detailed comparison.
Instead, we compare our results to the calculation in pionless EFT
\cite{Phillips:2008hn}. In this framework pions are integrated out and
$P$-odd interactions are fully described by contact interactions among
nucleons and, as such, the asymmetry in $pp$ scattering depends on
only one LEC, in the notation of Ref.~\cite{Phillips:2008hn},
$\mathcal A_{pp}$. 
The authors performed an analysis of the data points at $13.6$ and $45$ MeV and found $\mathcal A_{pp} = (1.3 \pm 0.3)\cdot 10^{-14}\,\mathrm{MeV}^{-3}$. In order to compare to the pionless approach we should set $h_\pi =0$. From Fig.~\ref{chi} we infer that this means $C = -(3.0 \pm 1.0)\cdot 10^{-6}$, in order to describe the data. Translating this to the notation of Ref.~\cite{Phillips:2008hn} we obtain a value for $\mathcal A_{pp}$
\begin{equation}
 \mathcal A_{pp} = -\frac{C}{2\Fp \Lambda_\chi^2} = (1.6\pm 0.5)\cdot 10^{-14}\,\mathrm{MeV}^{-3},
\end{equation}
in good agreement with the pionless result. 
Of course, non-vanishing values of $h_\pi$ can give very different values for $C$. 
Notice further that the above comparison should not be taken too
seriously since the LEC $C$ we consider in this work is, strictly
speaking, a bare quantity. On the other hand, the quoted value for
$\mathcal A_{pp}$ 
corresponds to a renormalized LEC at the scale  $\mu = m_\pi$. 

As mentioned, more experimental data is needed to further constrain
the LECs. If additional data is at odds with our allowed ranges for
$h_\pi$ and $C$, for example if upcoming experiments on the angular
asymmetry in $\vec n p\rightarrow d\gamma$ find a value of $h_\pi \sim
10^{-7}$ while, simultaneously, a new experiment on the $pp$ asymmetry
constrains $h_\pi \sim 10^{-6}$, it might be that higher-order
corrections to the $P$-odd potential need to be taken into account. In
fact, by analogy to the $P$-conserving case where next-to-next-to-leading order (N$^2$LO) and N$^3$LO contributions are very relevant, this might be expected. On the other hand, the analysis of Ref. \cite{KaiserPodd} shows that  certain corrections to the TPE-diagrams, which are important for the $P$-conserving potential, are small in the $P$-violating case. A full calculation of the N$^2$LO $P$-odd potential is necessary to say more about this potential issue.

In summary, we have investigated the longitudinal asymmetry in proton-proton scattering in chiral effective field theory. We calculated the asymmetry up to next-to-leading order in the parity-violating potential. By a careful comparison with the experimental data we have extracted allowed ranges for the two relevant parity-odd low-energy constants. The allowed ranges are consistent with theoretical calculations of the LECs and with experimental limits. However, more data is required in order to extract preciser values of the coupling constants.

\subsection*{Acknowledgements}

We thank Andreas Nogga for many helpful comments and discussions, Dieter Eversheim for providing helpful information about details of the Bonn experiment, and Matthias Schindler for clarifications of the pionless calculation.
This work is supported in part by the DFG and the NSFC through funds 
provided to the Sino-German CRC 110 ``Symmetries and the Emergence of 
Structure in QCD'', by the EU (HadronPhysics3) and ERC project 259218 NUCLEAREFT.

\appendix

\section{Partial-wave decomposition of the $P$-odd potential} 
\label{secpwd}

In order to solve the LS equation, it is necessary to have a partial wave 
decomposition of the potential. Details on the decomposition of the 
$P$-even potential can be found in Ref.~\cite{Epelbaum:2004fk} and here 
we consider the $P$-odd potential. Let us first ignore isospin, we then 
need to decompose a potential of the form
\begin{equation} 
V = i  (\vec \sigma_1\times \vec \sigma_2)\cdot \vec q\, f(q)~,
\end{equation}
where the form of $f(q)$ depends on whether we look at the TPE or the contact 
potential. Apart from isospin we have
\begin{eqnarray}\label{pwd}
V^{l^\prime l\, s^\prime s}_{j}(p^\prime, p) &=& \langle p^\prime\,(l^\prime
s^\prime) j || V || p\,(ls)j\rangle\nonumber\\
&=&-216\pi \sum_k \hat k^{3/2}(-1)^k g_k(q)\!\!\!\sum_{\lambda_1+\lambda_2=1} 
\sqrt{\frac{\hat \lambda_1\hat \lambda_2}{\hat \lambda_1! \hat \lambda_2!}} 
(p)^{\lambda_1} (-p^\prime)^{\lambda_2}\nonumber\\
&&\times C(\lambda_1\,k\,l ; 000)\, C(\lambda_2\,k\,l^\prime ; 000) 
\bma \lambda_1&\lambda_2&1 \\ k&k&0\\ l&l^\prime&1  \ema \nn\\
&&\times (-1)^l \sqrt{\hat s \hat {s^\prime}\hat j}\bma l^\prime&\l&1 \\ 
s^\prime&s&1\\ j&j&0  \ema \bma 1/2 &1/2&1 \\ 1/2&1/2&1\\ s^\prime&s&1  \ema~, 
\end{eqnarray}
where $\hat x = 2x+1$ and 
\begin{equation}
g_k(q) = \int_{-1}^{1} dx^\prime\, P_k(x^\prime) f(q(x^\prime))~, 
\end{equation} 
in terms of $q^2(x^\prime) = p^2 +p^{\prime\,2}-2 p p^\prime x^\prime$ 
and $P_k(x^\prime)$ are the Legendre polynomials.  To be specific, 
for the TPE potential
\[ f(q(x^\prime)) = \frac{1}{(2\pi)^3} V_{\text{TPE}}(q^\prime, \Lambda_S)e^{-p^6/\Lambda^6}e^{-p^{\prime\,6}/\Lambda^6}~.\]

To include isospin, we should multiply by the following factor
\begin{eqnarray}\label{pwdiso}
\langle(\frac{1}{2}\frac{1}{2}) t^\prime m_t^\prime | (\vec \tau_1+ 
\vec \tau_2)^3 | (\frac{1}{2} \frac{1}{2}) t m_t\rangle
&=& C(t\,1\,t^\prime;m_t\,0\,m_t^\prime)\langle(\frac{1}{2} \frac{1}{2}) 
t^\prime m_t^\prime || (\vec \tau_1+ 
\vec \tau_2)^3 || 
(\frac{1}{2}\frac{1}{2}) t m_t\rangle \nn\\
&=& C(t\,1\,t^\prime;m_t\,0\,m_t^\prime) 6\sqrt{\hat t}\,[1+(-1)^{t+t^\prime}]
\bma 1/2 &1/2&1 \\ 1/2&1/2&0\\ t^\prime&t&1  \ema.
\end{eqnarray}
In the case of proton-proton scattering we can put $t^\prime = t = m_t^\prime = m_t =1$. 

\section{Solution of the LS equation in momentum space} 
\label{LSsolve}

There are two main ways of approaching the problem in the sense that we can 
treat the $P$-odd potential either perturbatively or nonperturbatively. 
We begin with the latter approach. The first step involves the removal of the 
$i\epsilon$ in the numerator by writing
\[ \frac{1}{E- p^{\prime \prime\,2}/m_p + i\epsilon} 
= \frac{m_p}{q_0^2- p^{\prime \prime\,2} + i\epsilon} = 
\frac{m_p}{q_0 + p^{\prime \prime}} \left(\frac{\mathcal P}{q_0 - p^{\prime
      \prime}} - i\pi \delta(q_0 -p^{\prime \prime})\right)~,\]
where $\mathcal P$ denotes  the principal value integral, and $E = q_0^2/m_p$ 
such that $q_0$ is the on-shell momentum.
We can now write the LS equation as 
\begin{eqnarray}
T^{l^\prime l\, s^\prime s}_j (p^\prime, p, E) &=& V^{l^\prime l\, s^\prime  s}_j 
(p^\prime, p) +m_p \sum_{l^{\prime \prime}\,s^{\prime \prime}} \mathcal P\!\!
\int_0^{p_{\mathrm{max}}} 
dp^{\prime \prime}\, V^{l^\prime l^{\prime \prime}\, s^\prime s^{\prime
    \prime}}_j 
(p^\prime, p^{\prime \prime})\frac{p^{\prime \prime\,2}}{q_0^2
- p^{\prime \prime\,2} } T^{l^{\prime \prime} l\, s^{\prime \prime}
s}_j (p^{\prime \prime}, p, E)\nonumber\\
&& - i\frac{\pi m_p q_0}{2}\sum_{l^{\prime \prime}\,s^{\prime \prime}}  
V^{l^\prime l^{\prime \prime}\, s^\prime s^{\prime \prime}}_j (p^\prime, q_0)
T^{l^{\prime \prime} l\, s^{\prime \prime} s}_j (q_0, p, E)~,
\end{eqnarray}
where we introduced $p_{\mathrm{max}} \gg \Lambda$ which corresponds to 
the final grid point used in the numerical solution. 
We now subtract the divergence in the first integral and add it back 
again and write
\begin{eqnarray}
T^{l^\prime l\, s^\prime s}_j (p^\prime, p, E) 
&=& V^{l^\prime l\, s^\prime s}_j (p^\prime, p) \nonumber\\
&&+m_p \sum_{l^{\prime \prime}\,s^{\prime \prime}}  
\int_0^{p_{\mathrm{max}}} dp^{\prime \prime}\, 
\bigg[V^{l^\prime l^{\prime \prime}\, s^\prime s^{\prime \prime}}_j (p^\prime, 
p^{\prime \prime})\frac{p^{\prime \prime\,2}}{q_0^2- p^{\prime
      \prime\,2} } T^{l^{\prime \prime} l\, s^{\prime \prime} s}_j 
(p^{\prime \prime}, p, E)\nonumber\\
&& - V^{l^\prime l^{\prime \prime}\, s^\prime s^{\prime \prime}}_j 
(p^\prime, q_0)\frac{q_0^2}{q_0^2- p^{\prime \prime\,2} }
T^{l^{\prime \prime} l\, s^{\prime \prime} s}_j (q_0, p, E)\bigg]
\nonumber\\
&&+ m_p q_0^2 V^{l^\prime l^{\prime \prime}\, s^\prime s^{\prime \prime}}_j 
(p^\prime, q_0) T^{l^{\prime \prime} l\, s^{\prime \prime} s}_j (q_0, p, E) 
\mathcal P\!\!  \int_0^{p_{\mathrm{max}}} dp^{\prime \prime} \frac{1}{q_0^2- p^{\prime \prime\,2} }
\nonumber\\
&& - i\frac{\pi m_p q_0}{2}\sum_{l^{\prime \prime}\,s^{\prime \prime}}  
V^{l^\prime l^{\prime \prime}\, s^\prime s^{\prime \prime}}_j (p^\prime, q_0)
T^{l^{\prime \prime} l\, s^{\prime \prime} s}_j (q_0, p, E)~,
\end{eqnarray}
where the first integral has no pole so the principal value has been removed. 
The second integral can be done analytically and gives
\[ \mathcal P\!\!  \int_0^{p_{\mathrm{max}}} dp^{\prime \prime} 
\frac{1}{q_0^2- p^{\prime \prime\,2} } = \frac{1}{2q_0} 
\ln\frac{ p_{\mathrm{max}} + q_0}{ | p_{\mathrm{max}}-q_0|} ~.
\]

 The LS equation can now be solved numerically. The main difference with
 respect to only strong interactions is that more channels are coupled. 
Where in the limit of no parity violation (and isospin violation) there are 
two coupled and two uncoupled channels, in this case there are in general 
four coupled channels. In the case of $pp$ scattering there are 
always less channels (two coupled channels if $j=0$, one uncoupled channel 
if $j$ is odd, and three coupled channels if $j>0$ and even). In general, 
we solve the whole $T$-matrix at once. We write it as
 \begin{equation} \label{Tmat}
 T^{l^\prime l\, s^\prime s}_j = \bma T^{j-1\,j-1 \, 11}_j & T^{j-1\,j+1 \,
   11}_j&  T^{j-1\,j \, 11}_j& T^{j-1\,j \, 10}_j \\
 T^{j+1\,j-1 \, 11}_j & T^{j+1\,j+1 \, 11}_j&  T^{j+1\,j \, 11}_j& T^{j+1\,j \, 10}_j \\
 T^{j\,j-1 \, 11}_j & T^{j\,j+1 \, 11}_j&  T^{j\,j \, 11}_j& T^{j\,j \, 10}_j \\
 T^{j\,j-1 \, 01}_j & T^{j\,j+1 \, 01}_j&  T^{j\,j \, 01}_j& T^{j\,j \, 00}_j 
 \ema~. \end{equation}
 The top-left $2\times2$ matrix corresponds to the ``standard" coupled channels
 and the $33$ and $44$ entries are the ``standard" uncoupled channels. The
 entries connecting $j$ and $j\pm 1$ are zero in the absence of parity
 violation. The entries $T^{j\,j \, 10}_j$ and $T^{j\,j \, 01}_j$ remain zero
 unless there is isospin violation which changes total isospin in the strong interaction. 
 
 The other option is to solve the LS equation perturbatively. Ignoring all
 indices, the LS equation becomes
 \[ T = V + V G_0 T~,\]
 where $V = V_P + V_{\slashPsub}$, with $V_P$ denoting the $P$-conserving potential and $V_{\slashPsub}$ the 
$P$-violating potential. If we treat $V_{\slashPsub}$ as a perturbation we can use
first-order perturbation theory and write $T= T_P + T_{\slashPsub}$ as well. The leading 
equation becomes
  \[ T_P = V_P + V_P G_0 T_P~.\]
 This is just the ordinary strong LS equation which can be solved with the 
methods described above. The first-order equation becomes
  \[ T_{\slashPsub} = V_{\slashPsub} + V_P G_0 T_{\slashPsub} + V_{\slashPsub} G_0 T_P\qquad \rightarrow\qquad 
  (1 - V_P G_0)T_{\slashPsub} = V_{\slashPsub} + V_{\slashPsub} G_0 T_P~.\]
  The leading-order equation can be rewritten into
  \[1-V_P G_0 = (1+T_P G_0)^{-1}\]
  such that 
 \begin{equation} 
  T_{\slashPsub} = (1+T_P G_0)(V_{\slashPsub} + V_{\slashPsub} G_0 T_P) 
  = V_{\slashPsub} + V_{\slashPsub} G_0 T_P +  T_P G_0 V_{\slashPsub} + T_P G_0 V_{\slashPsub} G_0 T_P~, 
 \end{equation}
which can be solved directly. We have checked explicitly that the perturbative 
and the nonperturbative treatments give the same solution for the $T$-matrix, 
if the $P$-odd potential is small enough.


\begin{thebibliography}{99}

\bibitem{Desplanques:1979hn}
  B.~Desplanques, J.~F.~Donoghue and B.~R.~Holstein,
  Annals Phys.\  {\bf 124} (1980) 449.


\bibitem{Holsteinreview}
  W.~C.~Haxton and B.~R.~Holstein,
  Prog.\ Part.\ Nucl.\ Phys.\  {\bf 71} (2013) 185.


\bibitem{Schindler:2013yua}
  M.~R.~Schindler and R.~P.~Springer,
  Prog.\ Part.\ Nucl.\ Phys.\  {\bf 72} (2013) 1.  

\bibitem{Bernard:2006gx}
  V.~Bernard and U.-G.~Mei{\ss}ner,
  Ann.\ Rev.\ Nucl.\ Part.\ Sci.\  {\bf 57} (2007) 33.
  
\bibitem{Epelbaum:2008ga}
  E.~Epelbaum, H.-W.~Hammer and U.-G.~Mei{\ss}ner,
  Rev.\ Mod.\ Phys.\  {\bf 81} (2009) 1773.
 
\bibitem{Machleidt:2011zz}
  R.~Machleidt and D.~R.~Entem,
  Phys.\ Rept.\  {\bf 503} (2011) 1.

\bibitem{Kaplan:1992vj}
  D.~B.~Kaplan and M.~J.~Savage,
  Nucl.\ Phys.\ A {\bf 556} (1993) 653
   [Erratum-ibid.\ A {\bf 570} (1994) 833]
   [Erratum-ibid.\ A {\bf 580} (1994) 679].
   
\bibitem{Zhu}
  S.-L.~Zhu, C.~M.~Maekawa, B.~R.~Holstein, M.~J.~Ramsey-Musolf and U. van Kolck,
  Nucl.\ Phys.\ A {\bf 748} (2005) 435.
  
\bibitem{KaiserPodd}
  N.~Kaiser,
  Phys.\ Rev.\ C {\bf 76} (2007) 047001.    
  
\bibitem{Savage:1998rx}
  M.~J.~Savage and R.~P.~Springer,
  Nucl.\ Phys.\ A {\bf 644} (1998) 235
   [Erratum-ibid.\ A {\bf 657} (1999) 457].  

\bibitem{Savage:2000iv}
  M.~J.~Savage,
  Nucl.\ Phys.\ A {\bf 695} (2001) 365.  
  
 \bibitem{Girlanda:2008ts}
  L.~Girlanda,
  Phys.\ Rev.\ C {\bf 77} (2008) 067001. 

\bibitem{Phillips:2008hn}
  D.~R.~Phillips, M.~R.~Schindler and R.~P.~Springer,
  Nucl.\ Phys.\ A {\bf 822} (2009) 1.

\bibitem{Holstein:2009zzb}
  B.~R.~Holstein,
  Eur.\ Phys.\ J.\ A {\bf 41} (2009) 279.

\bibitem{Ordonez:1993tn}
  C.~Ordonez, L.~Ray and U.~van Kolck,
  Phys.\ Rev.\ Lett.\  {\bf 72} (1994) 1982.
  
  \bibitem{Ordonez:1993tn2}
  C.~Ordonez, L.~Ray and U.~van Kolck,
  Phys.\ Rev.\ C {\bf 53} (1996) 2086.
  

\bibitem{Epelbaum:1998ka}
  E.~Epelbaum, W.~Gl\"ockle and U.-G.~Mei{\ss}ner,
  Nucl.\ Phys.\ A {\bf 637} (1998) 107.
  
  \bibitem{Epelbaum:1998ka2}
  E.~Epelbaum, W.~Gl\"ockle and U.-G.~Mei{\ss}ner,
  Nucl.\ Phys.\ A {\bf 671} (2000) 295.  

\bibitem{Entem:2003ft}
  D.~R.~Entem and R.~Machleidt,
  Phys.\ Rev.\ C {\bf 68} (2003) 041001.
    
\bibitem{Epelbaum:2004fk}
  E.~Epelbaum, W.~Gl\"ockle and U.-G.~Mei{\ss}ner,
  Nucl.\ Phys.\ A {\bf 747} (2005) 362.  
    
  
  \bibitem{Epelbaum:2003gr}
  E.~Epelbaum, W.~Gl\"ockle and U.-G.~Mei{\ss}ner,
  Eur.\ Phys.\ J.\ A {\bf 19} (2004) 125.    

    
\bibitem{NDA}
A.V. Manohar and H. Georgi, 
Nucl. Phys. B {\bf 234} (1984)  189.

\bibitem{NDA2}
H. Georgi and L. Randall,
Nucl. Phys. B {\bf 276} (1986) 241.

\bibitem{Kaiser:1989fd}
N.~Kaiser and U.-G.~Mei{\ss}ner,
  Nucl.\ Phys.\ A {\bf 499} (1989) 699.

\bibitem{Meissner:1998pu}
  U.-G.~Mei{\ss}ner and H.~Weigel,
  Phys.\ Lett.\ B {\bf 447} (1999) 1.
  
\bibitem{hpilatt}
J. Wasem,
Phys. Rev.  C {\bf 85} (2012) 022501.

\bibitem{Adelberger:1983zz}
  E.~G.~Adelberger, M.~M.~Hindi, C.~D.~Hoyle, H.~E.~Swanson, R.~D.~Von Lintig and W.~C.~Haxton,
  Phys.\ Rev.\ C {\bf 27} (1983) 2833.
  
  \bibitem{Adelberger:1983zz2}
  S.~A.~Page, H.~C.~Evans, G.~T.~Ewan, S.~P.~Kwan, J.~R.~Leslie, J.~D.~Macarthur, W.~Mclatchie and P.~Skensved {\it et al.},
  Phys.\ Rev.\ C {\bf 35} (1987) 1119.
  
\bibitem{Haxton:1981sf}
  W.~C.~Haxton,
  Phys.\ Rev.\ Lett.\  {\bf 46} (1981) 698.
  
  \bibitem{Liu:2006dm}
  C.-P.~Liu,
  Phys.\ Rev.\ C {\bf 75} (2007) 065501.
  
  \bibitem{Partanen:2012qw}
  T.~M.~Partanen, J.~A.~Niskanen and M.~J.~Iqbal,
  Eur.\ Phys.\ J.\ A {\bf 48} (2012) 119.
  
  
  \bibitem{Vincent:1974zz}
  C.~M.~Vincent and S.~C.~Phatak,
  Phys.\ Rev.\ C {\bf 10} (1974) 391.

\bibitem{Walzl:2000cx}
  M.~Walzl, U.-G.~Mei{\ss}ner and E.~Epelbaum,
  Nucl.\ Phys.\ A {\bf 693} (2001) 663.
  
\bibitem{Carlson:2001ma}
  J.~Carlson, R.~Schiavilla, V.~R.~Brown and B.~F.~Gibson,
  Phys.\ Rev.\ C {\bf 65} (2002) 035502.
  
\bibitem{Kong:1999sf}
  X.~Kong and F.~Ravndal,
  Nucl.\ Phys.\ A {\bf 665} (2000) 137.
    
\bibitem{Taylor}
 J.~R.~Taylor, Scattering Theory, (Dover Publications, 2006).

  
\bibitem{Driscoll:1988hg}
  D.~E.~Driscoll and G.~A.~Miller,
  Phys.\ Rev.\ C {\bf 39} (1989) 1951. 
  

  

\bibitem{Eversheim:1991tg}
  P.~D.~Eversheim, W.~Schmitt, S.~E.~Kuhn, F.~Hinterberger, P.~von Rossen, J.~Chlebek, R.~Gebel and U.~Lahr {\it et al.},
  Phys.\ Lett.\ B {\bf 256} (1991) 11.
  \bibitem{Eversheim:1991tg2}
  P.~D.~Eversheim, private communication. 
  

\bibitem{Kistryn:1987tq}
  S.~Kistryn, J.~Lang, J.~Liechti, T.~Maier, R.~Muller, F.~Nessi-Tedaldi, M.~Simonius and J.~Smyrski {\it et al.},
  Phys.\ Rev.\ Lett.\  {\bf 58} (1987) 1616.  
  

\bibitem{Berdoz:2001nu}
  A.~R.~Berdoz {\it et al.}  [TRIUMF E497 Collaboration],
  Phys.\ Rev.\ Lett.\  {\bf 87} (2001) 272301.  

 \bibitem{Driscoll:1989jv}
  D.~E.~Driscoll and U.-G.~Mei{\ss}ner,
  Phys.\ Rev.\ C {\bf 41} (1990) 1303. 

\bibitem{Stoks:1994wp}
  V.~G.~J.~Stoks, R.~A.~M.~Klomp, C.~P.~F.~Terheggen and J.~J.~de Swart,
  Phys.\ Rev.\ C {\bf 49} (1994) 2950.  
  
\bibitem{Kaplan:1998xi}
  D.~B.~Kaplan, M.~J.~Savage, R.~P.~Springer and M.~B.~Wise,
  Phys.\ Lett.\ B {\bf 449} (1999) 1.
  
\bibitem{Epelbaum:2001fm}
  E.~Epelbaum, U.-G.~Mei{\ss}ner, W.~Gl\"ockle and C.~Elster,
  Phys.\ Rev.\ C {\bf 65} (2002) 044001.  

\bibitem{Berengut:2013nh}
  J.~C.~Berengut, E.~Epelbaum, V.~V.~Flambaum, C.~Hanhart, U.-G.~Mei{\ss}ner, J.~Nebreda and J.~R.~Pelaez,
  Phys.\ Rev.\ D {\bf 87} (2013) 085018.  

  
  

  \end{thebibliography}
\end{document}